\documentclass[]{emulateapj}

\usepackage{graphicx,amssymb,verbatim,natbib}

\newcommand\hi{\hbox{H$\rm I$~}}

\newcommand\cii{\hbox{C$\rm II$~}}
\newcommand\oisiii{\hbox{O$\rm I$/Si$\rm II$~}}
\newcommand\se{Sect.~}
\newcommand\lya{Ly$\alpha$}
\newcommand\gp{$g_{475}$}
\newcommand\wt{$w(\theta)$}
\newcommand\rp{$r_{625}$}
\newcommand\ip{$i_{775}$}
\newcommand\inp{$I_{814}$}
\newcommand\zp{$z_{850}$}
\newcommand\bp{$B_{435}$}
\newcommand\up{$U_{300}$}
\newcommand\vp{$V_{606}$}
\newcommand\ks{$K_{S}$}

\shorttitle{Galaxies in a protocluster at $z=4.1$}
\shortauthors{Overzier et al.}

\begin{document}      
                      
\title{Lyman break galaxies, Ly$\alpha$ emitters and a radio galaxy in a protocluster at $z=4.1$\altaffilmark{1,2}.}

\author{Roderik A. Overzier\altaffilmark{3,7}, R.J. Bouwens\altaffilmark{4}, N.J.G. Cross\altaffilmark{5},  B.P. Venemans\altaffilmark{6}, G. K. Miley\altaffilmark{3}, A.W. Zirm\altaffilmark{7}, N. Ben\'{\i}tez\altaffilmark{8}, J.P. Blakeslee\altaffilmark{9}, D. Coe\altaffilmark{8}, R. Demarco\altaffilmark{7}, H.C. Ford\altaffilmark{7}, N.L. Homeier\altaffilmark{7}, G.D. Illingworth\altaffilmark{4}, J.D. Kurk\altaffilmark{10}, A.R. Martel\altaffilmark{7}, S. Mei\altaffilmark{11}, I. Oliveira\altaffilmark{3,12}, H.J.A. R\"ottgering\altaffilmark{3}, Z.I. Tsvetanov\altaffilmark{7,13}, W. Zheng\altaffilmark{7}}

\email{overzier@pha.jhu.edu}

\altaffiltext{1}{Based on observations made with the NASA/ESA Hubble
Space Telescope, which is operated by the Association of Universities
for Research in Astronomy, Inc., under NASA contract NAS 5-26555. 
These observations are associated with program \#9291.}
\altaffiltext{2}{Based on observations carried out at the European Southern Observatory, Paranal, Chile, programs 071.A-0495(A) and 073.A-0286(A).}
\altaffiltext{3}{Leiden Observatory, Postbus 9513, 2300 RA Leiden, Netherlands}
\altaffiltext{4}{UCO/Lick Observatory, University of California, Santa Cruz, CA 95064}
\altaffiltext{5}{Royal Observatory Edinburgh, Blackford Hill, Edinburgh, EH9 3HJ, UK}
\altaffiltext{6}{Institute of Astronomy, Madingley Road, Cambridge CB3 0HA, UK}
\altaffiltext{7}{Department of Physics and Astronomy, The Johns Hopkins University, 3400 North Charles Street, Baltimore, MD 21218}
\altaffiltext{8}{Inst. Astrof\'{\i}sica de Andaluc\'{\i}a (CSIC), Camino Bajo de Hu\'{e}tor, 24, Granada 18008, Spain}
\altaffiltext{9}{Department of Physics and Astronomy, Washington State University, Pullman, WA 99164-2814}
\altaffiltext{10}{Max-Planck-Institut fur Astronomie, Koenigstuhl 17, D-69117, Heidelberg, Germany}
\altaffiltext{11}{GEPI, Observatoire de Paris, Section de Meudon, Meudon Cedex, France}
\altaffiltext{12}{Division of Geological and Planetary Sciences, California Institute of Technology, Pasadena, CA 91125, USA}
\altaffiltext{13}{National Aeronautics and Space Administration, Washington, DC}

\begin{abstract}
  We present deep {\it HST/ACS} observations in \gp\rp\ip\zp\ towards
  the $z=4.1$ radio galaxy TN J1338--1942 and its overdensity of $>$30
  spectroscopically confirmed \lya\ emitters (LAEs).  We select 66
  \gp-band dropouts to \zp$_{,5\sigma}=27$, 6 of which are also a LAE.
  Although our color-color selection results in a relatively broad
  redshift range centered on $z=4.1$, the field of TN J1338--1942 is
  richer than the average field at the $>$5$\sigma$ significance,
  based on a comparison with GOODS.  The angular distribution is
  filamentary with about half of the objects clustered near the radio
  galaxy, and a small, excess signal ($2\sigma$) in the projected pair
  counts at separations of $\theta<10\arcsec$ is interpreted as being
  due to physical pairs.  The LAEs are young (a few $\times10^7$ yr),
  small ($\langle r_{hl}\rangle=0\farcs13$) galaxies, and we derive a
  mean stellar mass of $\sim10^{8-9}$ M$_\odot$ based on a stacked
  \ks-band image.  We determine star formation rates, sizes,
  morphologies, and color-magnitude relations of the \gp-dropouts and
  find no evidence for a difference between galaxies near TN
  J1338--1942 and in the field.  We conclude that environmental trends
  as observed in clusters at much lower redshift are either not yet
  present, or are washed out by the relatively broad selection in
  redshift.  The large galaxy overdensity, its corresponding mass
  overdensity and the sub-clustering at the approximate redshift of TN
  J1338--1942 suggest the assemblage of a $>10^{14}$ M$_\odot$
  structure, confirming that it is possible to find and study cluster
  progenitors in the linear regime at $z\gtrsim4$.
\end{abstract}



\section{Introduction}
\label{sec:intro}

\noindent
The formation and evolution of structure in the Universe is a
fundamental field of research in cosmology. Clusters of galaxies
represent the most extreme deviation from initial conditions in the
Universe, and are therefore good evolutionary probes for studying the
formation of the large-scale structure. While clusters of galaxies
have been studied extensively in the relatively nearby Universe, their
evolutionary history becomes obscure beyond roughly half the Hubble
time \citep{blakeslee06,mullis05,stanford06}.  Their progenitors are
extremely difficult to identify when the density contrast between the
forming cluster and the field becomes small, and mass condensations on
the scales of clusters are extremely rare at any epoch
\citep{kaiser84}.

Even the most distant clusters known contain, besides a population of
star-forming galaxies, an older population of relatively red and
massive galaxies \citep[e.g.][]{dressler99,vandokkum00,goto05}.  The
scatter in the color-magnitude relation for cluster ellipticals at
$z\sim1$ is virtually indistinguishable from that at low redshift,
suggesting that some of the galaxy populations in clusters have very
old stellar populations
\citep[e.g.][]{stanford98,blakeslee03_clusters,wuyts04,holden05}.  The
epoch of cluster and cluster galaxy formation is presumed to be marked
by the violent build-up of the stellar mass and morphologies of these
early-type galaxies.

Large samples of star-forming Lyman break galaxies (LBGs) that are
selected in the rest-frame UV have been used to probe the cosmic star
formation rate history as well as the large-scale structure far beyond
$z=1$.  The epoch corresponding to the peak in the star formation rate
density ($z\sim2-3$) is preceded by a significant but modest decrease
in the star formation rate density from $z\sim3$ out to $z\sim6$
\citep{madau96,steidel99,giavalisco04_results,bouwens07}.  Deep
near-IR NICMOS observations in and around the GOODS fields find small
samples of $z\sim7-8$ galaxies
\citep{bouwensillingworth06,bouwens04_z7} and indicate that this
decline in the number of $UV$ luminous galaxies continues from
$z\sim7-8$ \citep{bouwensillingworth06}.  This suggests that cosmic
star formation is a rather extended process, with only modest but
continual changes over time.

LBGs, as well as the partially overlapping population of \lya\
emitters (LAEs), are strongly clustered at $z=3-5$, and are highly
biased relative to predictions for the dark matter distribution
\citep{giavalisco98,adelberger98,ouchi04_r0,lee05,kashikawa05}.  The
biasing becomes stronger for galaxies with higher rest-frame UV
luminosity \citep{giavalisco01}.  \citet{ouchi04_r0} found that the
bias may also increase with redshift and dust extinction. They suggest
that the reddest LBGs could be connected with sub-mm sources or the
extremely red objects \citep[EROs,][]{elston88,mccarthy01,daddi02}.
By comparing the number densities of LBGs to that of dark halos
predicted by \citet{sheth99}, they concluded that LBGs at $z\sim4$ are
hosted by halos of $1\times10^{11}-5\times10^{12}$ M$_\odot$, and that
the descendants of those halos at $z=0$ have masses that are
comparable to the masses of groups and clusters. The derived halo
occupation numbers of LBGs increase with luminosity from a few tenths
to roughly unity, implying that there is only one-to-one
correspondence between halos and LBGs at the highest masses.

Studies of sizes indicate that high-redshift star-forming galaxies are
compact in size ($\sim0.1$\arcsec--0.3\arcsec), while large
($\gtrsim0.4$\arcsec) low surface brightness galaxies are rare
\citep{bouwens04_sizes}, and there is a clear decrease in size with
redshift for objects of fixed luminosity
\citep{ferguson04,bouwens04_sizes}. Morphological analysis of LBGs
indicates that they often possess brighter nuclei and more disturbed
profiles than local Hubble types degraded to the same image quality
\citep[e.g.][]{lotz04}.

Despite these advances in the study of the evolution of the highest
redshift galaxies, {\it galaxy clusters} have been studied out to only
$z=1.5$ while it is clear that they began forming at much earlier
epochs.  Finding and studying these cluster progenitors may yield
powerful tests for (semi-)analytical models and $N$-body simulations
of structure formation, and give new clues to how the most massive
structures in the Universe came about.

Several good candidates for galaxy overdensities, possibly
`protoclusters\footnote{The term {\it protocluster} is commonly used
  to describe non-virialized galaxy overdensities at high redshift
  ($z\gtrsim2$) that, when evolved to the present day, have estimated
  masses comparable to those of the virialized galaxy clusters. See
  Sect. 6 for a more physical definition.}', have been discovered at
very high redshift
\citep[e.g.][]{pascarelle96,keel99,francis01,moller01,steidel98,steidel05,shimasaku03,ouchi05,kashikawa07}.
These structures have been found often as by-products of wide field
surveys using broad or narrow band imaging.

Luminous radio galaxies have also been used to search for overdense
regions within the large-scale structure at high redshift.  The
association of distant, powerful radio galaxies with massive galaxy
and cluster formation is mainly based on two observational clues.
First, some high redshift radio galaxies are very massive
systems\footnote{We note that the measure of the absolute stellar
  masses of galaxies is still a matter of debate given the different
  implementations in the literature of the post-AGB stellar phase in
  the modeling of the rest frame near-infrared galaxy emission
  \citep[e.g., see][]{maraston06,bruzual07,eminian07}.}
\citep[e.g.][]{debreuck02,dey97,pentericci01,villarmartin05,miley06,seymour07},
suggesting that their host galaxies are likely to be the progenitors
of massive red sequence galaxies or even the brightest cluster
galaxies (BCGs) that dominate the deep potential wells of clusters.
Second, in some cases high redshift radio galaxies have been shown to
have companion galaxies. At $1.5<z<2$, there is evidence for
overdensities of red galaxies associated with radio sources,
consistent with moderately rich Abell-type clusters
\citep[e.g.][]{sanchez99,sanchez02,thompson00,hall01,best03,wold03}.
At $z>2$, several radio galaxies have been found to be surrounded by
overdensities of LAEs of a few, discovered through deep narrow band
imaging and spectroscopic follow-up with the Very Large Telescope
(VLT) of the European Southern Observatory
\citep[e.g.][]{pentericci00,kurk03,venemans02,venemans04,venemans05,venemans07}.

Focussing on the excesses of LAEs discovered in the vicinity of
distant radio sources, we are performing a survey of LBGs in such
radio-selected protoclusters with the {\it Advanced Camera for
  Surveys} on the {\it Hubble Space Telescope}
\citep[HST/ACS;][]{ford98}.  In \citet{miley04} and \citet{overzier06}
we reported on the detection of a significant population of LBGs in
the (projected) region around radio galaxies at $z=4.1$ (TN
J1338--1942) and $z=5.2$ (TN J0924--2201). Here, we will present a
detailed analysis of the ACS observations of protocluster TN
J1338--1942 at $z=4.1$, augmented by ground-based observations with
the VLT. This structure is amongst the handful of overdense regions so
far discovered at $z>4$, as evidenced by 37 LAEs that represent a
surface overdensity of $\sim5$ compared to other fields
\citep[][]{venemans02,venemans07}. The FWHM of the velocity
distribution of the LAEs is 625 km s$^{-1}$. The mass overdensity as
well as the velocity structure is consistent with the global
properties of $z\sim4$ protoclusters derived from simulations
\citep[e.g.][]{delucia04,suwa06}.  The radio galaxy itself is highly
luminous in the rest-frame UV, optical and the sub-mm, indicating a
high star formation rate. It has a complex morphology which we have
interpreted as arising from AGN feedback on the forming ISM and a
massive starburst-driven wind \citep{zirm05}.

The main issues that we will attempt to address here are the
following. What are the star formation histories, physical sizes and
morphologies of LAEs and LBGs?  In particular we wish to study these
properties in relation to the overdense environment that the TN
J1338--1942 field is believed to be associated with, analogous to
galaxy environmental dependencies observed at lower redshifts and
predicted by models
\citep[e.g.][]{kauffmann04,postman04,delucia05}. How does
sub-clustering associated with the TN J1338--1942 structure compare to
the `field'? What is the mass overdensity of the structure, and what
is the relation to lower redshift galaxy clusters?  In Sect. 2 we will
describe the observations, data reduction and methods. We present our
sample of LBGs in Sect. 3, and describe the rest-frame UV and optical
properties of LBGs and LAEs. In Sect. 4 we present the results of a
nonparametric morphological analysis.  In Sect. 5 we will present
further evidence for a galaxy overdensity associated with TN
J1338--1942 and investigate its clustering properties. We conclude
with a summary and discussion of our main results (Sect. 6).  We use a
cosmology in which $H_0=72$ km s$^{-1}$ Mpc$^{-1}$, $\Omega_M=0.27$,
and $\Omega_\Lambda=0.73$ \citep{spergel03}.  The luminosity distance
is 37.1 Gpc and the angular scale size is 6.9 kpc arcsec$^{-1}$ at
$z=4.1$.  The lookback time is 11.9 Gyr, corresponding to an epoch
when the Universe was approximately 11\% of its current age.  All
colors and magnitudes quoted in this paper are expressed in the $AB$
system \citep{oke71}.


\section{Observations and data reduction} 

\subsection{ACS imaging}

\begin{figure*}[t]
\begin{center}
\includegraphics[width=0.7\textwidth,height=5cm]{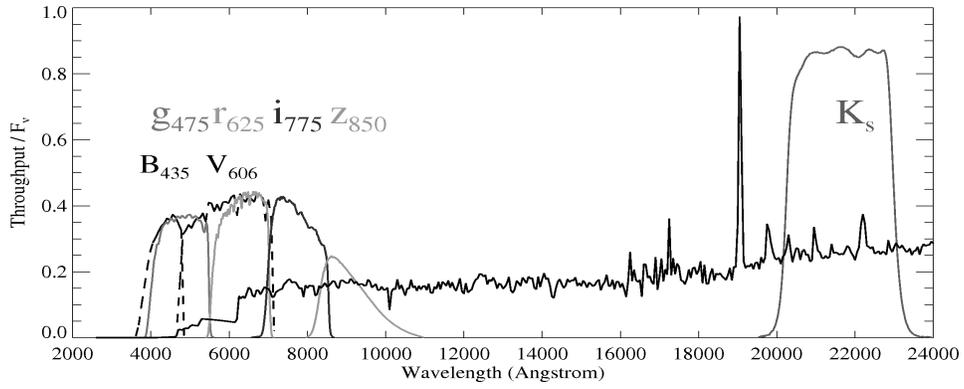}
\end{center}
\caption{\label{fig:filters} Total effective throughput of the HST/ACS
  \gp\rp\ip\zp\ and VLT/ISAAC $K_S$ filters. The galaxy spectrum shown
  is the $SB2$ template from \citet{benitez00} redshifted to $z=4.1$,
  and applying the attenuation prescription of \citet{madau96}.  The
  GOODS \bp\ and \vp\ filters are indicated by dashed lines for
  comparison.}
\end{figure*}

\noindent
We observed one field with the ACS around the radio galaxy TN
J1338--1942 (henceforward `TN1338').  These observations were part of
the ACS Guaranteed Time Observing high redshift cluster program. To
search for candidate cluster members on the basis of a Lyman-break at
the approximate wavelength of \lya\ redshifted to $z=4.1$, we used the
Wide Field Channel to obtain imaging through the broadband
filters\footnote{We use \gp, \rp, \ip\ and \zp\ to denote magnitudes
  in the HST/ACS passbands F475W, F625W, F775W and F850LP,
  respectively, or to denote the passbands themselves.} \gp, \rp, \ip,
and \zp\ (Fig. \ref{fig:filters}).  The total observing time of 18
orbits was split into 9400 s in each of \gp\ and \rp, and 11700 s in
each of \ip\ and \zp.

\begin{figure*}[t]
\begin{center}
\includegraphics[width=0.7\textwidth]{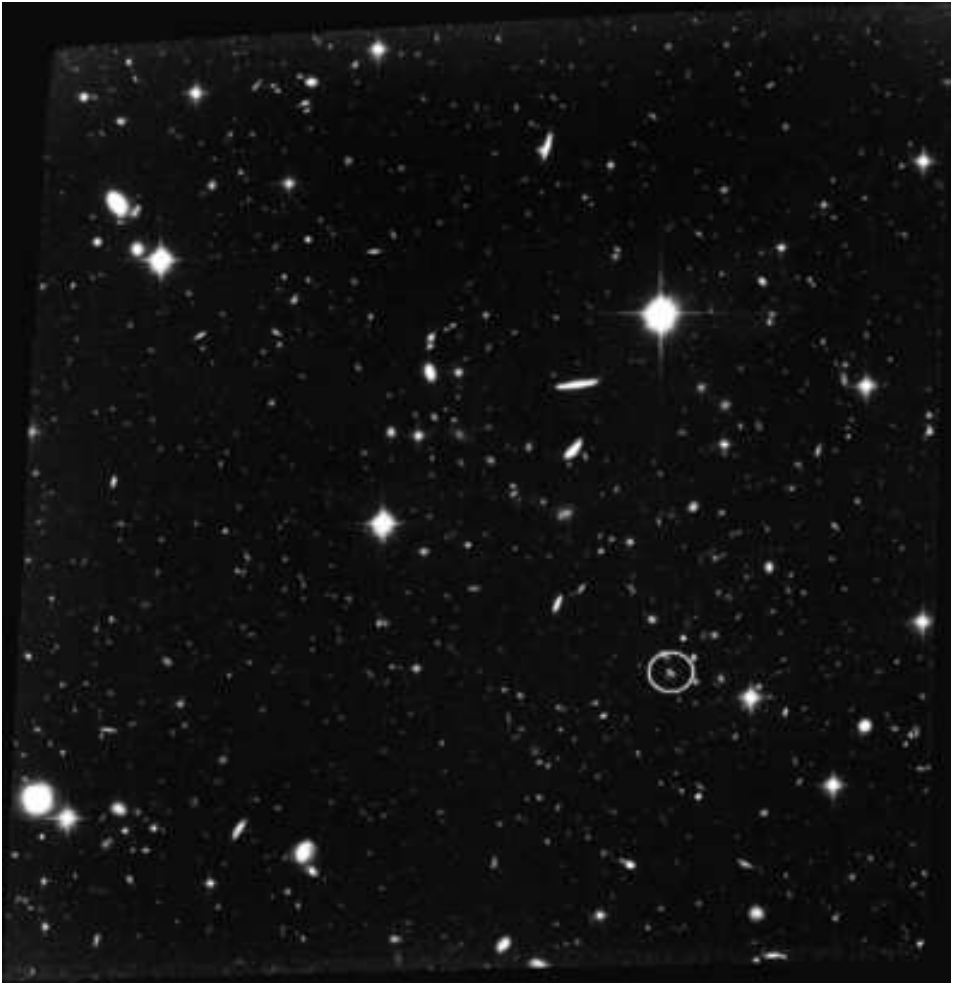}
\end{center}
\caption{\label{fig:colorfield}The ACS field showing \gp\ in blue,
  \rp\ in green and \zp\ in red. The field measures 11.7 arcmin$^2$.
  The radio galaxy TN1338 stands out clearly as the bright green
  object (white circle).  }
\end{figure*}

Each orbit of observation time was split into two 1200 s exposures to
facilitate the removal of cosmic rays. The data were reduced using the
ACS pipeline science investigation software \citep[{\it
  Apsis};][]{blakeslee03}. After initial processing of the raw data
through CALACS at STScI (bias/dark subtraction and flat-fielding), the
following processing was performed by {\it Apsis}: empirical
determination of image offsets and rotations using a triangle matching
algorithm, background subtraction, the rejection of cosmic rays and
the geometric correction and combining of exposures through drizzling
using the STSDAS Dither package. The final science images have a scale
of $0\farcs05$ pixel$^{-1}$. The total field of view is 11.7
arcmin$^2$. The radio galaxy ($\alpha_{J2000}=13^h38^m30^s$,
$\delta_{J2000}=-19\degr42\arcmin30\arcsec$) is located about
$1\arcmin$ away from the image centre.  The field further includes 12
of the 37 spectroscopically confirmed LAEs
\citep{venemans02,venemans07}.  The resultant color image of the field
is shown in Fig. \ref{fig:colorfield}, with \gp\ in blue, \rp\ in
green, and \zp\ in red.  The radio galaxy clearly stands out as the
sole `green' object in the entire field, due to its prominent halo of
\lya\ emission observed in \rp\ \citep[see][]{venemans02,zirm05}.

We used the ACS zeropoints from \citet[][]{sirianni05}, and an
extinction value of $E(B-V)=0.096$ mag from \citet{schlegel98}. We
measured the limiting magnitudes from the RMS of noise fluctuations in
10\,000 square apertures of varying size that were distributed over
the images in regions free of objects.  Details of the observations
are given in Table \ref{tab:log}.

\subsection{VLT optical spectroscopy and NIR imaging}

\noindent
We obtained 10 hours of VLT/FORS2 spectroscopy in service
mode\footnote{Program ID: 071.A-0495(A)}.  The instrumental setup, the
seeing conditions, and the method of processing of the data were
similar as described in \citet[][]{venemans02,venemans05}.

Near-infrared data in the $K_S$-band were obtained with
VLT/ISAAC\footnote{Program ID: 073.A-0286(A)}.  We observed a
$2\farcm4\times2\farcm4$ field for 2.1 hours in March 2002, and for
5.4 hours in a partly overlapping field in 2004.  After dark
subtraction, flat fielding and rejection of science frames of poor
quality, the data for each night was individually processed into a
combined image using the XDIMSUM package in IRAF\footnotetext{IRAF is
  distributed by the National Optical Astronomy Observatories, which
  are operated by the Association of Universities for Research in
  Astronomy, Inc., under cooperative agreement with the National
  Science Foundation.}. Since only the data taken on the night of
March 26 2002 was considered photometric, the combined images of the
other nights were scaled to match that particular night using several
unsaturated stars for reference. We derived the zeropoint based on
observations of the near-IR photometric standard FS 142.  However, we
had to adjust the zeropoint by 0.2 magnitudes to match the magnitudes
of several 2MASS stars in the field.  The seeing was $\sim$0\farcs5
(FWHM), and the galactic extinction in $K_S$ was 0.036 mag.

Next, each of the combined images with the native ISAAC scale of
0\farcs148 pixel$^{-1}$ was corrected for geometric distortion by
projecting it onto the 0\farcs05 pixel$^{-1}$ ACS \ip\ image using the
tasks $\mathtt{GEOMAP/GEOTRAN}$ in $\mathtt{IRAF}$.  The projection
had a typical accuracy of $\sim$1.5 ACS pixels (RMS).  The registered
images were combined using a weighting based on the variance measured
in a source-free region of each image.  The limiting $2\sigma$ depth
in the AB\footnote{$K_{s,AB}=K_{s,Vega}+1.86$} system was 25.2
magnitudes for a circular aperture of 1\farcs4 diameter. Areas that
are only covered by either the 2002 or the 2004 data are shallower by
0.5 and 0.3 mag, respectively.  The $K_S$-band data cover 81\% of the
ACS field, and contain the radio galaxy and 11 LAEs.

\subsection{Object detection and photometry}
\label{sec:detect}

\noindent
Object detection and photometry was done using the
$\mathtt{SExtractor}$ software package of \citet{bertin96}.  We used
SExtractor in double-image mode, where object detection and aperture
determination are carried out on the so-called ``detection image'',
and the photometry is carried out on the individual filter images. For
the detection image we used an inverse variance weighted average of
the \rp, \ip\ and \zp\ images, and a map of the total exposure time
per pixel was used as the detection weight map. Photometric errors
were calculated using the root mean square (RMS) images from {\it
  Apsis}.  These images contain the absolute error per pixel for each
output science image. We detected objects by requiring a minimum of 5
connected pixels at a threshold of 1.5 times the local background
($S/N$ of $>3.35$). The values for SExtractor's deblending parameters
($\mathtt{DEBLEND\_MINCONT=0.1}$, $\mathtt{DEBLEND\_NTHRESH=8}$) were
chosen to limit the extent to which our often clumpy $z\sim4$
\gp-dropouts were split into multiple objects. Our `raw' detection
catalog contained 3994 objects.  We rejected all objects which had
$S/N$ less than 5 in \zp, where we define $S/N$ as the ratio of counts
in the isophotal aperture to the errors on the counts. The remaining
2022 objects were considered real objects, although they still contain
a small fraction of artefacts and objects that were split up.

We used SExtractor's $\mathtt{MAG\_AUTO}$ to estimate total magnitudes
within an aperture radius of $2.5\times r_{\mathrm{Kron}}$
\citep{kron80}. However, when accurate color estimation is more
important than estimating a galaxy's total flux, for example in the
case of color-selection or when determining photometric redshifts,
isophotal magnitudes are preferred because of the higher $S/N$ and the
smaller contribution of neighboring sources. Therefore we calculate
{\it galaxy colors} from isophotal magnitudes. These procedures are
optimal for (faint) object detection and aperture photometry with ACS
\citep{benitez04}.

Optical-NIR (observed-frame) colors were derived from combining the
ACS data with lower-resolution groundbased $K_S$ data in the following
way. We used $\mathtt{PSFMATCH}$ in IRAF to determine the 2D kernel
that matches the point spread function (PSF) in the ACS images to that
obtained in the $K_S$-band, and convolved the ACS images with this
kernel. The photometry was done using SExtractor in double image mode,
using the $K_S$-band image for object detection.  Colors involving the
NIR data were determined in circular apertures with a diameter of
1\farcs4, although we used a 3\farcs0 diameter aperture for the large
radio galaxy.

\subsection{Aperture and completeness corrections}

\noindent
The photometric properties of galaxies are usually measured using
source extraction algorithms such as SExtractor. We can conveniently
use this software to determine aperture corrections and completeness
limits as a function of e.g., the `intrinsic' or real apparent
magnitude, half-light radius ($r_{hl}$) or the shape of the galaxy
surface brightness profile \citep[see
also][]{benitez04,giavalisco04_survey}. To this end we populated the
ACS \zp\ image with artificial galaxies consisting of a 50/50 mix of
exponential and de Vaucouleurs profiles. We simulated $\sim10\,000$
galaxies with $\sim200$ per simulated image to avoid over-crowding. We
took uniformly distributed half-light radii in the range
0\farcs1--1\farcs0, and uniformly distributed axial ratios in the
range 0.1--1.0. Galaxies were placed on the images with random
position angles on the sky.  Using the zeropoint we scaled the counts
of each galaxy to uniformly populate the range \zp$=20-28$ mag. We
added Poisson noise to the simulated profiles, and convolved with the
\zp\ PSF. SExtractor was used to recover the model galaxies as
described in Sect. \ref{sec:detect}.

\begin{figure*}[t]
\begin{center}
\includegraphics[width=\textwidth]{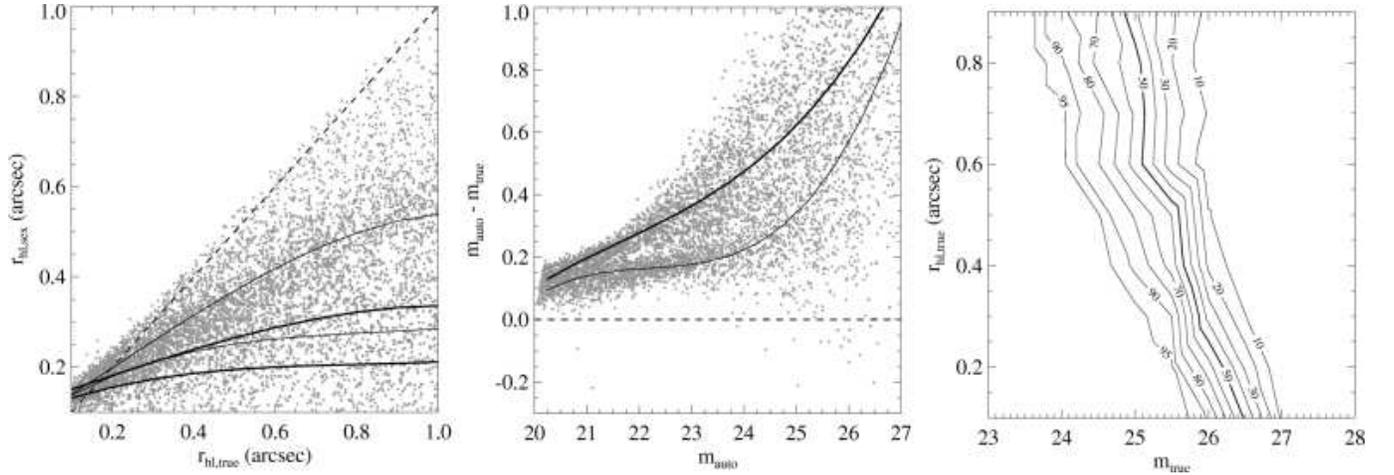}
\end{center}
\caption{\label{fig:completeness}{\it Left:} Intrinsic $r_{hl,z}$
  versus $r_{hl,z}$ measured by SExtractor for de Vaucouleurs profiles
  (thick lines) and exponentials (thin lines). The difference between
  intrinsic and measured radius is smaller for exponentials. The
  intrinsic sizes are increasingly underestimated when going to
  fainter magnitudes, e.g. from \zp$\sim24$ mag (top lines) to
  \zp$\sim26$ mag (bottom lines). -- {\it Middle:} The difference in
  \zp\ between MAG\_AUTO and total `intrinsic' magnitudes for de
  Vaucouleurs profiles (thick line) and exponentials (thin line).
  {\it Right:} Completeness limits in \zp\ as a function of total
  `intrinsic' magnitudes and $r_{hl}$, where completeness is defined
  as the ratio of the number of objects detected to the number of
  artificial objects added to the image (50\% exponential; 50\% de
  Vaucouleurs). The 50\% completeness limit lies at an intrinsic
  magnitude of $\approx26.5$ for small sources. When expressed in
  terms of MAG\_AUTO, the completeness limits shown here should be
  some $\sim$0.5--1.0 mag fainter, due to an underestimate of the
  `total' flux for faint galaxies.}
\end{figure*}

Approximately 75\% of the artificial galaxies were detected. In
Fig. \ref{fig:completeness} (left panel) we show the measured $r_{hl}$
versus the input `intrinsic' $r_{hl}$.  Radii are increasingly
underestimated as the input radii become larger, because the surface
brightness gets fainter as $r_{hl}^2$.  On average, the radius is
underestimated by about 50\% for a \zp$\sim26$ mag object with an
intrinsic half-light radius of 0\farcs4.  The discrepancy between
input and output radius is generally smaller for an exponential than
for a de Vaucouleurs profile.  In Fig. \ref{fig:completeness} (middle
panel) we show the aperture corrections defined by the difference
between MAG\_AUTO and the total simulated magnitude. The amount of
flux missed rises significantly towards fainter magnitudes, with a
0.5--1.0 mag correction for objects with output magnitudes of
\zp$=25-27$ mag.  Finally, in Fig. \ref{fig:completeness} (right
panel) we show the \zp\ completeness limits as a function of \zp\ and
$r_{hl}$.  About 50\% completeness is reached at \zp$=26-26.5$ mag for
unresolved or slightly resolved sources. Note that the 50\%
completeness limit will lie at measured MAG\_AUTO magnitudes that are
fainter by 0.5--1.0 mag, given the significant aperture corrections
shown in the middle panel of Fig. \ref{fig:completeness}.

Throughout the paper, we apply approximate corrections to the physical
quantities derived from measured $r_{hl}$ and magnitudes
(e.g. physical sizes, luminosities, and SFRs) based on the above
results for exponential profiles. Angular sizes and magnitudes quoted
are always as measured.

\subsection{Photometric redshift technique}

\noindent
We used the Bayesian Photometric Redshift code (BPZ) of
\citet{benitez00} to estimate galaxy redshifts, $z_B$. For a complete
description of BPZ and the robustness of its results, we refer the
reader to \citet{benitez00} and \citet{benitez04}. Our library of
galaxy spectra is based on the elliptical, intermediate ($Sbc$) and
late type spiral ($Scd$), and irregular templates of
\citet{coleman80}, augmented by two starburst galaxy templates with
$E(B-V)\sim0.3$ ($SB2$) and $E(B-V)\sim0.45$ ($SB3$) from
\citet{kinney96}, and two simple stellar population (SSP) models with
ages of 5 Myr and 25 Myr from \citet{bruzualcharlot03}. As shown by
\citet{coe05}, the latter two templates improve the accuracy of BPZ
for very blue, young high redshift galaxies in the Hubble Ultra Deep
Field \citep[UDF,][]{beckwith06}.  BPZ uses a parameter `ODDS' defined
as $P(|z-z_B|<\Delta z)$ that gives the total probability that the
true redshift is within an uncertainty $\Delta z$. For the uncertainty
we can take the empirical accuracy of BPZ for the HDF-N which has
$\sigma=0.06(1+z_B)$. For a Gaussian probability distribution a
$2\sigma$ confidence interval centered on $z_B$ would get an ODDS of
$>0.95$.  The empirical accuracy of BPZ is $\sigma\approx0.1(1+z_B)$
for objects with $I_{814}\lesssim24$ and $z\lesssim4$ observed in the
\bp\vp\inp-bands with ACS to a depth comparable to our observations
\citep{benitez04}. Note that we will be applying BPZ to generally
fainter objects at $z\sim4$ observed in \gp\rp\ip\zp. The true
accuracy for such a sample has yet to be determined empirically.  The
accuracy of BPZ may be improved by using certain {\it priors}. We
apply the commonly used magnitude prior that is based on the magnitude
distribution of galaxies in real observations (e.g. the HDF).

\subsection{Template-based color-color selection of protocluster LBG candidates}
\label{sec:select}

\noindent
We extracted LBGs from our catalogs using color criteria that are
optimized for detecting star-forming galaxies at $z\sim4$
\citep{steidel99,ouchi04_lf,giavalisco04_results}. To define the
optimal selection for our choice of filters we followed the approach
employed by \citet{madau96}. We used the evolutionary stellar
population synthesis model code GALAXEV (Bruzual \& Charlot 2003) to
simulate a large variety of galaxy spectral energy distributions
(SEDs) using: (i) the Padova 1994 simple stellar population model with
a \citet{salpeter55} initial mass function with lower and upper mass
cutoffs $m_L=0.1~M_\odot$ and $m_U=100~M_\odot$ of three metallicities
($0.2Z_\odot,0.4Z_\odot,Z_\odot$), and (ii) the predefined star
formation histories for instantaneous burst, exponentially declining
($\tau=0.01$ Gyr) and constant ($t=0.1,1.0$ Gyr) star formation.  We
extracted spectra with ages between 1 Myr and 13 Gyr, applied the
reddening law of \citet{calzetti00} with $E(B-V)$ of 0.0--0.5 mag, and
redshifted each spectrum to redshifts between 0.001 and 6.0, including
the effects of attenuation by the intergalactic medium (IGM) using the
\citet{madau96} recipe.  Galaxies were required to be younger than the
age of the Universe at their redshift, but other parameters were not
tied to redshift.  While this approach is rather simplistic due to the
fact that the model spectra are not directly tied to real observed
spectra and luminosity functions, it is reasonable to expect that they
at least span the range of allowed physical spectra. The resulting
library can then be used to define a set of color criteria for
selecting star-forming galaxies at the appropriate redshift, and
estimating color-completeness and contamination \citep{madau96}.

\begin{figure}[t]
\begin{center}
\includegraphics[width=0.6\columnwidth]{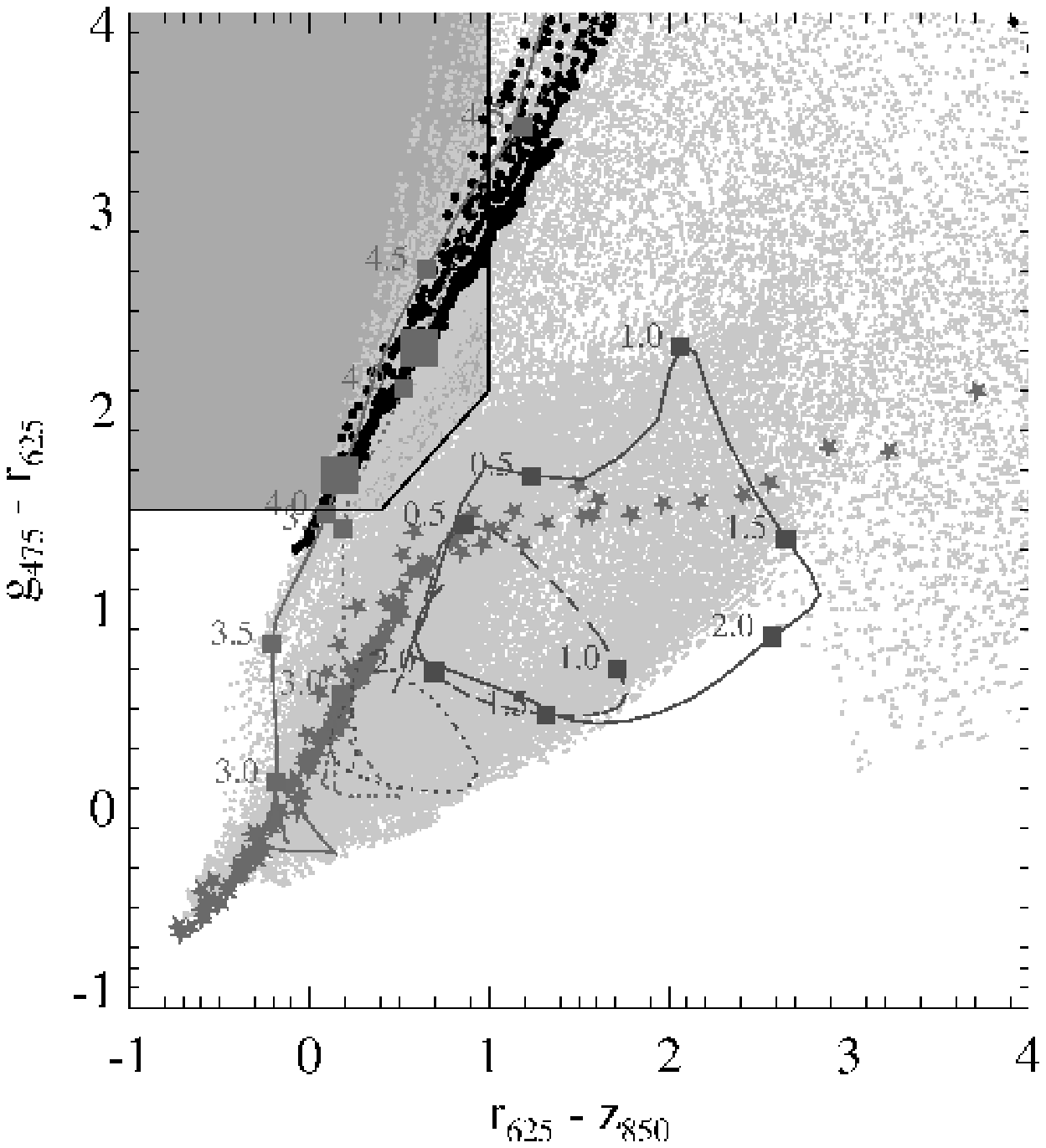}
\end{center}
\caption{\label{fig:grid1}\gp--\rp\ versus \rp--\zp\ for model 
  simulations using the \citet{bruzualcharlot03} libraries. Grey points
  indicate the modeled colors of a large set of SEDs in the redshift
  range $0<z<6$ (see text for details).  Galaxies at $z=4.1$ with ages
  less than 500 Myr are shown as large solid circles. Our selection criteria for selecting \gp-dropouts at $z\approx4.1$ 
  as defined by \gp--\rp$\ge1.5$, \gp--\rp$\ge$\rp--\zp$+1.1$,
  \rp--\zp$\le1.0$ are indicated by the shaded area. The spectral tracks are an elliptical (red solid
  line), an Sbc (red dashed line), an Scd (red dotted line), and a 100
  Myr constant star formation model with $E(B-V)=0.0$ mag (blue solid
  line) and $E(B-V)=0.2$ mag (blue dotted line). Redshifts are indicated
  along the tracks. The redshift of the overdensity of
  \citet{venemans02} is marked by the large blue squares
  ($z\approx4.1$). Stars (green) mark the stellar locus based on the
  stellar library of \citet{pickles98}.}
\end{figure}

We extracted the model colors by folding each spectrum through the
corresponding ACS filter transmission curves. No photometric scatter
was applied to the models.  The \gp--\rp\ and \rp--\zp\ color-color
diagram used to isolate LBGs at $z\sim4$ is shown in
Fig. \ref{fig:grid1}.  We elected to use the \rp--\zp\ color in
defining our selection region \citep[instead of the \rp--\ip\ color
used in][]{miley04} due to the greater leverage in wavelength.

\begin{figure}[t]
\begin{center}
\includegraphics[width=\columnwidth]{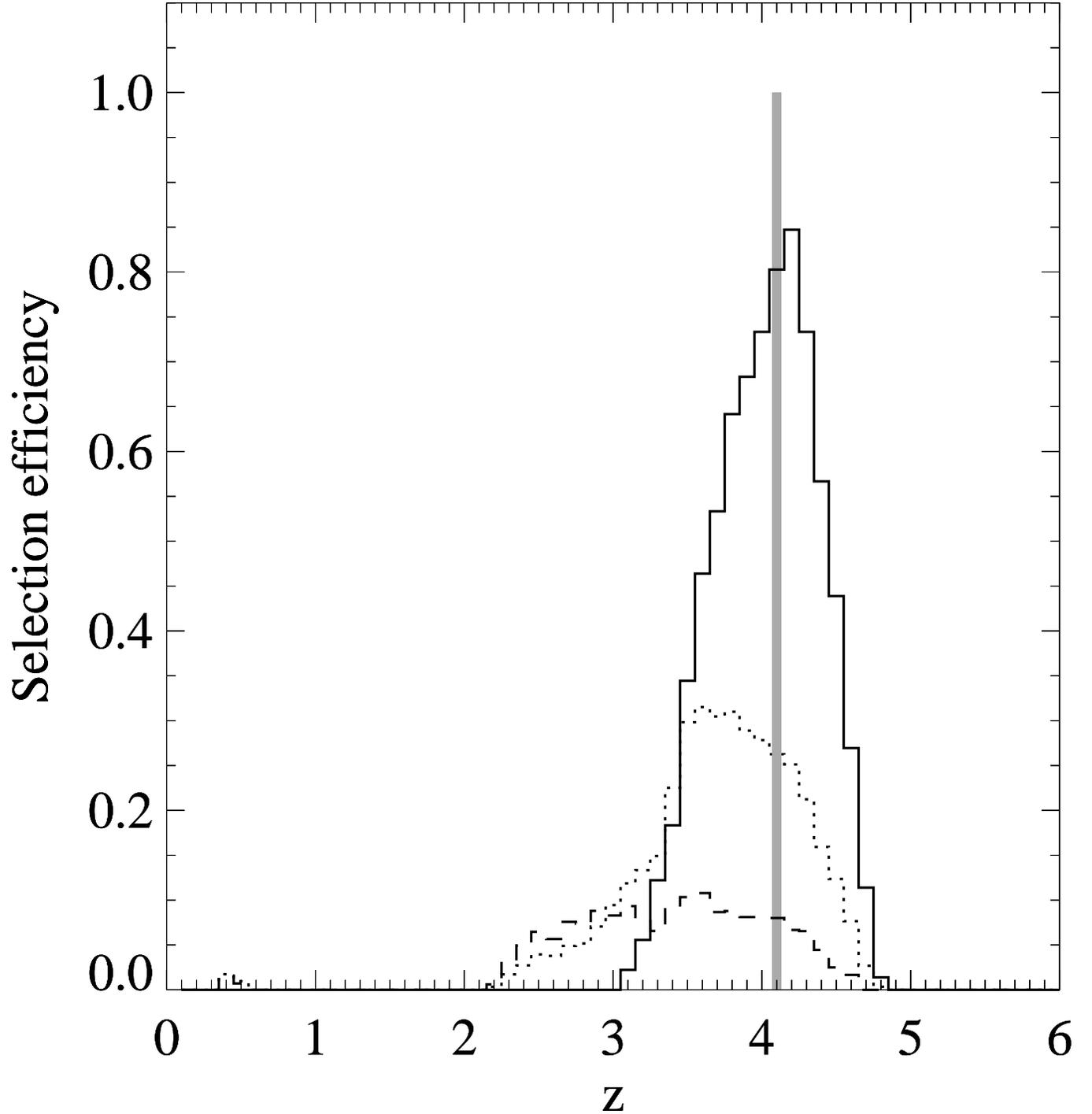}
\end{center}
\caption{\label{fig:grid2}Selection efficiency for $z\sim4$ LBGs.  The
  dotted histogram shows the fraction of model galaxies that meets the
  selection criteria in each redshift bin. The solid histogram shows
  the selection efficiency for model galaxies with ages less than 100
  Myr and $0<E(B-V)<0.3$. The dashed histogram shows the fraction of
  models with ages greater than 0.5 Gyr selected, illustrating
  possible contamination of our $z\sim4.1$ sample by relatively old
  galaxies at $z\sim2.5$. A (minor) source of contamination is the
  possible inclusion of Balmer-break objects at $z\sim0.5$. The shaded
  region indicates the redshift interval ($z=4.07-4.13$) of the
  protocluster defined by the radio galaxy and the Ly$\alpha$ emitters
  from \citet{venemans02}.}
\end{figure}

The color-color region that we use to select $z\sim4.1$ LBGs is defined as:
\begin{eqnarray}
\label{eq:criteria}
&&g_{475}-r_{625}\ge1.5,\nonumber \\
&&g_{475}-r_{625}\ge r_{625}-z_{850} + 1.1,\nonumber \\
&&r_{625}-z_{850}\le1.0.
\end{eqnarray}
Fig. \ref{fig:grid2} shows the color selection efficiency as a
function of redshift, defined as the number of galaxies selected in a
redshift bin, divided by the total number of model galaxies in that
redshift bin.  The dotted histogram indicates the fraction of model
galaxies meeting the selection criteria. The resulting redshift
distribution has an approximately constant maximum efficiency of
$\sim30$\% for $3.5<z<4.5$. If we limit the model galaxies to ages
less than 100 Myr and $0<E(B-V)<0.3$ (consistent with the average LBG
population at $z\sim3-4$ \citep{papovich01,steidel99}), the color
completeness (solid histogram) becomes $\sim90$\% for models at
$z\sim4.1$. The dashed histogram shows the fraction of models with
ages greater than 0.5 Gyr selected, illustrating the main sources of
contamination in our $z\sim4.1$ sample, namely from relatively old
galaxies at $z\sim2.5$ and the possible inclusion of Balmer-break
objects at $z\sim0.5$.

\subsection{GOODS simulated images}
\label{sec:sims}

\subsubsection{The simulations}
\label{sec:sims1}

\noindent
To determine whether TN1338 is also host to an overdensity of LBGs at
$z\sim4.1$, we will want to compare the number of $g_{475}$-dropouts
found in our ACS field with that found in a random field on the sky.
Unfortunately, at present, there are not many ACS fields available,
with comparable depths in $g_{475}$, $r_{625}$, and $z_{850}$ to carry
out such comparison.  We therefore avail ourselves of the four-band
GOODS field to make these comparisons.  The 3 orbit $B_{435}$, 2.5
orbit $V_{606}$, 2.5 orbit $i_{775}$, and 5 orbit $z_{850}$ coverage
is similar in depth and much larger in coverage, to the
$g_{475}r_{625}i_{775}z_{850}$ imaging we have on TN1338, suggesting
that with simple wavelength interpolation, we should be able to mirror
our TN1338 selection. 

Though there are many ways to have performed this interpolation, we
chose to perform the interpolation directly on the ACS data itself,
changing it from the observed $B_{435}V_{606}i_{775}z_{850}$ filter
set to the $g_{475}r_{625}i_{775}z_{850}$ filter set.  This
transformation was performed on a pixel-by-pixel basis, using the
formula:
\begin{small}
\begin{eqnarray}
f_{i,j}^Y &=& I_{i,j} ~ \textrm{g}(SED,Y,i_{775},z) ~+\nonumber\\
& & \Sigma_{X=X_L}^{X_H} \left ( \frac{|\lambda (Y) - \lambda (X)|}{\lambda (X_H) - \lambda(X_L)} \right ) ~\times~\nonumber\\
& &\textrm{g}(SED,Y,X,z) \Delta f_{i,j}^X,
\label{eqn:transform}
\end{eqnarray}
\end{small}
\noindent
where $f_{i,j}^Y$ is the flux at pixel $(i,j)$ in some band $Y$,
$I_{i,j}$ is the best-fit flux in each pixel (expressed as an
$i_{775}$-band flux), $\textrm{g}(SED,Y,X,z)$ is a generalized
$k$-correction from some band $X$ to another band $Y$ for some $SED$
and redshift $z$, and $\lambda (X)$ is the mean wavelength for some
band $X$. The summation $\Sigma_{X=X_L}^{X_H}$ runs over those bands
which immediately straddle the $Y$ band, and the $\Delta f_{i,j}^{X}$
terms account for the error in the fits to individual pixels.  The
best-fit fluxes $I_{i,j}$ were determined by minimizing
\begin{small}
\begin{equation}
\chi ^2 = \Sigma_X \left[ \frac{I_{i,j}\textrm{g}(SED,X,i_{775},z) - f_{i,j}^X}{\sigma_{i,j}^X} \right],
\end{equation}
\end{small}
\noindent
where $f_{i,j}^X$ and $\sigma_{i,j}^X$ are the flux and its
uncertainty, respectively, in the X band at pixel position $(i,j)$.
The error terms $\Delta f_{i,j}^X$ are equal to
$f_{i,j}-\textrm{g}(SED,X,i_{775},z)$.  The first term in
Eq.~\ref{eqn:transform} is a generalized $k$-correction applied to the
best-fit model SEDs, while the second is an interpolation applied to
the flux residuals from the fit.  This is nearly identical to
expressions from Appendix B1 of \citet{bouwens03} and represents a
slight update to that procedure.

The redshift, $z$, and spectral energy distribution, $SED$, that we
use for individual pixels are based upon an initial object catalog we
made of each field before doing the transformation.  Objects are
detected off a $\chi^2$ image \citep{szalay99} constructed from the
$V_{606}i_{775}z_{850}$-band using a fairly aggressive $3\sigma$
threshold and splitting parameter (SExtractor DEBLEND\_MINCONT=0.005).
Best-fit redshifts and SEDs are then estimated for each object from
the photometry.  These model parameters, in turn, are assigned to all
the pixels which make up these objects (according to the SExtractor
deblending maps), and thus used in the transformation given by
Eq.~\ref{eqn:transform}. Only objects having colors of
$(B_{435}-V_{606})>0.8$, $(B_{435}-V_{606})>0.6(V_{606}-z_{850})+0.5$,
$(B_{435}-V_{606})>3.375(V_{606}-z_{850})-4.575$ were
transformed. These colour limits were derived from the standard
\bp-dropout selection region (see Fig. 8 in Bouwens et al. 2007), but
are less restrictive in order to account for the fact that objects
which drop out of the \gp-band should have a slightly higher mean
redshift than objects which drop out of the \bp-band. Therefore, we
need to expand the size of our selection window to select galaxies at
slightly higher mean redshifts on average, i.e.  by taking galaxies
with a redder \vp--\zp\ color. Because objects with higher redshift
will also drop out of the \vp-band sooner, we have to decrease the
colour cut in \bp--\vp\ as well because the limit we can set on the
\bp--\vp\ colours will be weaker. The selection region thus includes
all objects expected in a standard \bp-dropout selection, but was
expanded to allow for objects at slightly higher and lower redshifts
that are potentially important when transforming the images to the
TN1338 filter set.

Since our ACS reduction of the TN1338 field had a different pixel
scale (i.e., 0\farcs05) than that of the GOODS v1.0 reduction
\citep[0\farcs03:][]{giavalisco04_survey}, we did not use that
reduction as the basis for our simulation of the CDF-S GOODS field.
Instead, we made use of an independent reduction we had made of the
GOODS field with {\it Apsis}.  That reduction was performed on a
0\farcs05 grid, using a procedure nearly identical to that described
in \citet{bouwens05_z6}, but using a `Lanzcos3' kernel (which matches
the TN1338 ACS reduction).

\subsubsection{Reliability of the simulations}
\label{sec:simtest}

\noindent
The image transformation method described in the previous section
provides us with a convenient way of converting the \bp\vp\ip\zp
imaging data available in the GOODS field into the \gp\rp\ip\zp\ bands
used for the TN1338 $g$-dropout selection.  In the limit of infinite
S/N data set and perfect model SEDs, the results obtained from this
transformation should be close to perfect.  However, in the real world
with finite S/N, it becomes necessary for us to test this method to
see how well it works in practice.  We do this by generating two
different simulations of the same field and then repeating the same
\gp-dropout selection on each simulation.  In the first simulation, we
generate the \gp\rp\ip\zp\ image set directly from some input model,
and in the second simulation, we first use this same model to generate
a \bp\vp\ip\zp\ image set (to mimic the GOODS data in both the depth
and passband coverage) and then convert this image set to the
\gp\rp\ip\zp\ bands using the image transformation method described in
the previous section.  By comparing the \gp-dropout selections
obtained from both methods, we can examine the effect that this
transformation method has on our $g$-dropout selection.  The same
input catalog is used for both simulations and was generated using the
\citet[][]{bouwens07} $z\sim4$ LF and a mean $UV$-continuum slope of
$-1.5$ with $1\sigma$ scatter of 0.6.  The profiles for sources in
this catalog were taken from real sources in the \citet[][]{bouwens07}
HUDF $B$-dropout sample.  The redshift of the input objects for the
simulation ranged from $z=3$ to $z=5$ and the limiting magnitude was
\zp$=27.5$.

\begin{figure}[t]
\begin{center}
\includegraphics[width=\columnwidth]{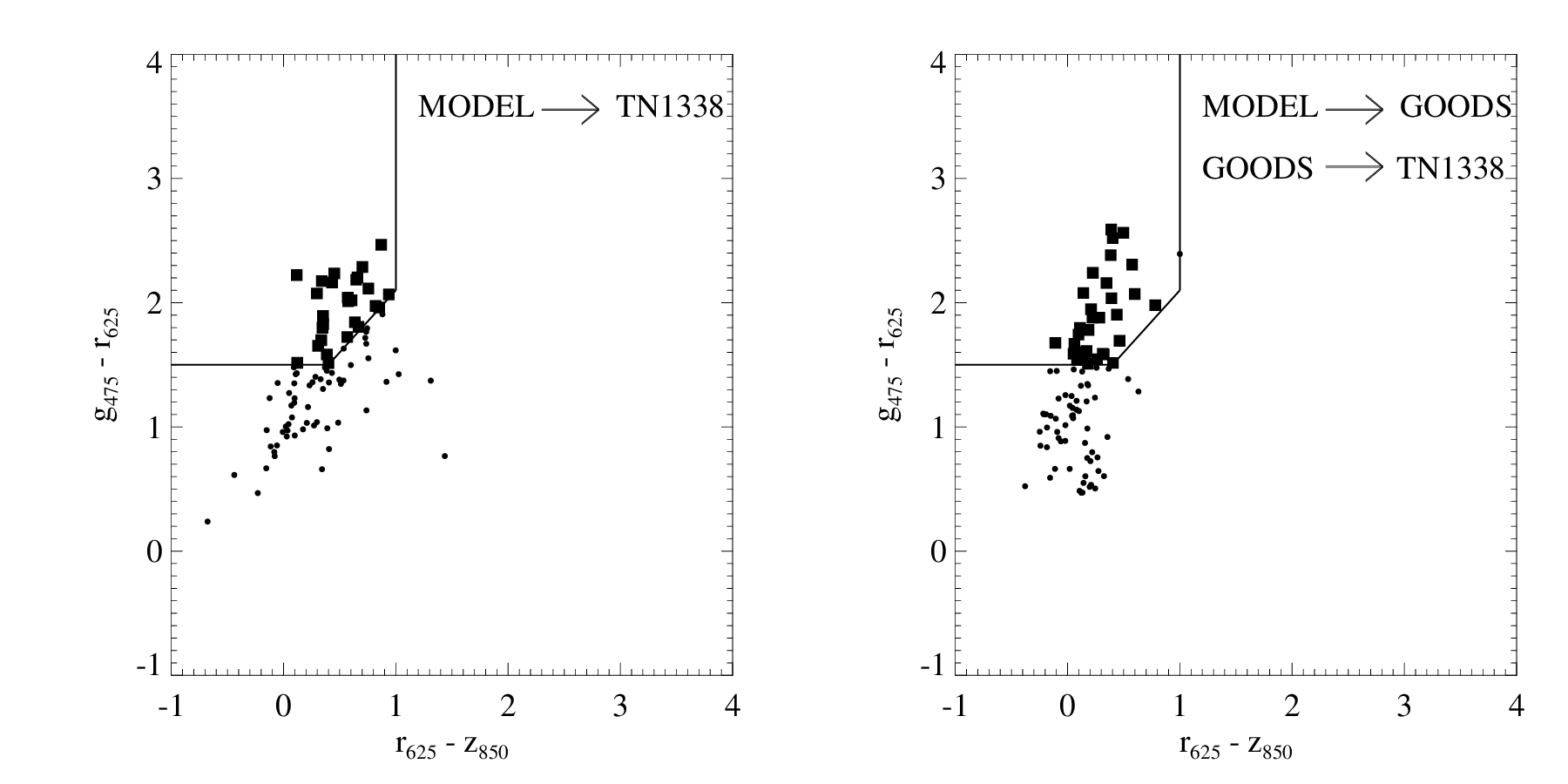}
\end{center}
\caption{\label{fig:clonetest}Colors and \gp-dropout samples derived
  from two different simulations of the TN1338 \gp\rp\ip\zp\ data set.
  In the first simulation ({\it left panel}), the input model was
  transformed directly in terms of the TN1338 \gp\rp\ip\zp filter
  set. In the second ({\it right panel}), the model was first
  transformed in the terms of the GOODS \bp\vp\ip\zp filter set and
  then transformed to the TN1338 \gp\rp\ip\zp\ filter set using the
  transformation method laid out in Section \ref{sec:sims}.  Objects
  detected in the images are indicated by small circles. Objects
  falling within our \gp-dropout selection window (marked by thick
  lines) are indicated by the large squares. Note that the
  transformation from model to GOODS to TN1338 ({\it right panel})
  introduces some small changes in the color distributions with
  respect to the transformation directly from model to TN1338 ({\it
    left panel}). This is due to the fact that in the right panel
  object detection and color selection are performed {\it twice}, and
  because of uncertainties introduced by determining the photometric
  redshifts of the objects in the simulated GOODS images. However, the
  \gp-dropout selections obtained from the images generated from our
  transformation method are nearly identical in number to those
  obtained from direct simulations of the same data set.  This
  suggests that it is reasonable to use the \gp-dropout samples
  obtained from the transformed images for quantifying the expected
  number densities (see Section \ref{sec:overdense}).}
\end{figure}

Objects were extracted from the two mock TN1338 datasets as described
in Sect. \ref{sec:detect}, and we selected a \gp-dropout sample from
each of the two datasets using the selection criteria detailed in
Sect. \ref{sec:select}. In Fig. \ref{fig:clonetest} we compare the
color-color diagram of objects detected in both the direct simulations
and those obtained from our image transformation method, respectively.
Objects (not) qualifying as \gp-dropouts as defined in
Sect. \ref{sec:select} are indicated by (small) large symbols. In
order to make a fair comparison, we have excluded objects at
$z\lesssim3.5$ and $z\gtrsim4.5$ in the left panel, because such
objects are not present in the right panel due to the wide \bp-dropout
color selection that is applied when transforming the GOODS images
into the TN1338 images (see Sect. \ref{sec:sims1}). Note that the
transformation from model to GOODS to TN1338 ({\it right panel})
introduces some small changes in the color distributions with respect
to the transformation directly from model to TN1338 ({\it left
  panel}). This is due to the fact that in the right panel object
detection and color selection are performed {\it twice}, and because
of uncertainties introduced by determining the photometric redshifts
of the objects in the simulated GOODS images. However, using different
limiting \zp\ magnitudes of [26.5,26.0,25.5] the ratio of the number
of \gp-dropouts detected/selected in our direct simulations to that in
our transformed GOODS images was [0.84,1.17,1.13]. At the faint end,
our image transformation method (comparable to the method used for
transforming the real GOODS dataset into a TN1338 dataset) thus
slightly overproduces the actual number of \gp-dropouts expected based
on the UDF input model. At the brighter limiting magnitudes, our image
transformation method slightly underproduces the number expected from
the direct method (at \zp$<25.5$ the difference of 13\% represents a
difference of just one object).  Overall, we conclude that the
\gp-dropout selections performed on the transformed images to be
fairly similar (within $\sim20\%$) in general in terms of the overall
numbers to that found on data obtained directly in those bands.  Thus,
we are confident in comparing the \gp-dropout counts found in the
transformed GOODS images to that found in the real TN1338 data (see
Sect. \ref{sec:overdense}).

\section{Results}

\noindent
In this section we apply the color-color section to the TN1338 field
to select a sample of candidate $z\sim4.1$ LBGs (\gp-dropouts) and
study their properties in Sect. \ref{sec:4.1}. In Sect.  \ref{sec:4.2}
we will study the same properties for the sample of $z=4.1$ LAEs.

\subsection{The \gp-dropout sample}
\label{sec:4.1}

\begin{figure}[t]
\begin{center}
\includegraphics[width=0.6\columnwidth]{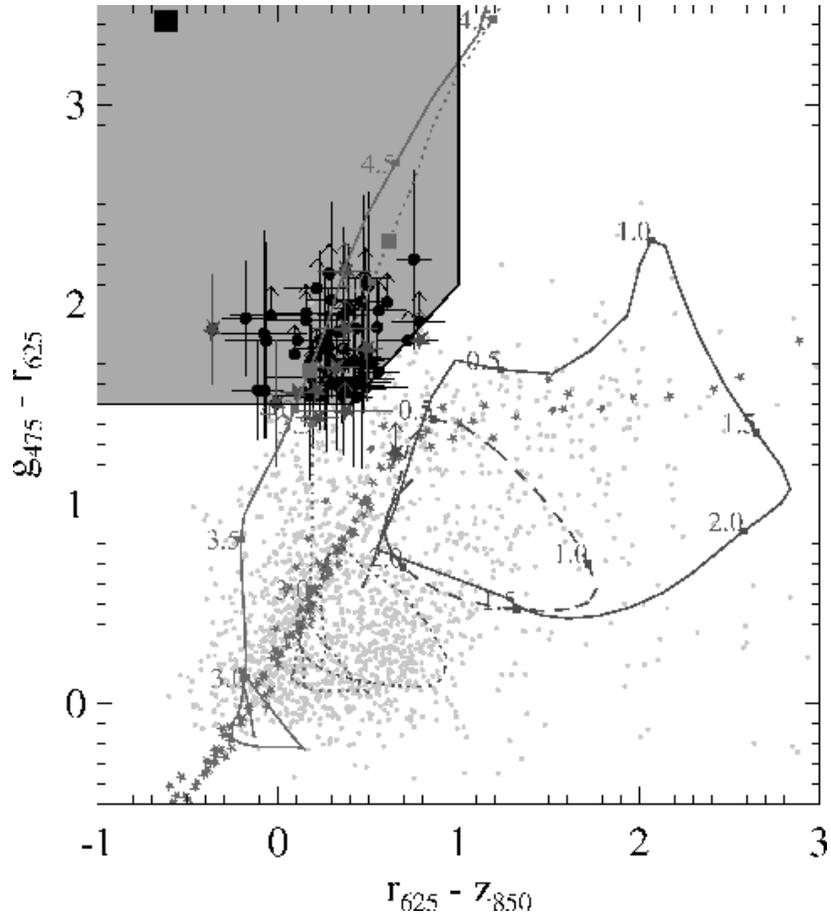}
\end{center}
\caption{\label{fig:cc}Color-color diagram of \gp-dropouts in TN1338
  (large circles) and the detection catalog (grey points).  The shaded
  region shows our selection window (Eq.\ 1). Confirmed LAEs from
  \citet{venemans02} are marked by red stars, and the radio galaxy by
  the large black square. Small green stars mark the stellar locus
  based on the stellar SED library of \citet{pickles98}.  See the
  caption of Fig. \ref{fig:grid1} for further details.}
\end{figure}

\noindent
Using the selection criteria defined in Eq. \ref{eq:criteria} we
extracted a sample of \gp-dropouts.  Although the stellar locus
\citep{pickles98} lies outside the region defined by our selection
criteria, we additionally required objects to have a SExtractor
stellarity index of $<0.85$ (non-stellar objects with high
confidence). Our final sample consisted of 66 objects with \zp$<$27.0
mag, 51 of which have \zp$<$26.5 mag, and 32 of which have \zp$<$26.0
mag. The color-color diagram is shown in Fig. \ref{fig:cc}.

\subsubsection{Star formation rates}

\noindent
The characteristic luminosity, $L^*_{z=4}$, of the LBG luminosity
function at $z\sim4$ corresponds to \zp$\sim25.0$
\citep{steidel99}. The sample contains two objects, one of which is
the radio galaxy, with a luminosity of $\sim6L^*$
(\ip$\approx23$). The remainder of the sample spans luminosities in
the range $\sim0.4-2L^*$, where we have applied aperture corrections
of up to $\sim1$ magnitude based on the exponential profiles in
Fig. \ref{fig:completeness}.

We calculated star formation rates (SFRs) from the emission-line free
UV flux at 1500 \AA\ (\ip) using the conversion between luminosity and
SFR for a Salpeter initial mass function (IMF) given in
\citet{madau98}: SFR (M$_\odot$ yr$^{-1}$) $=L_{1500 \textrm{\AA}}$
(erg s$^{-1}$ Hz$^{-1}$)$/8\times10^{27}$.  For ages that are larger
than the average time that late-O/early-B stars spend on the main
sequence, the UV luminosity is proportional to the SFR, relatively
independent of the prior star formation history.  The SFRs are listed
in Table \ref{tab:lbgs}. The radio galaxy and object \#367 each have a
SFR of $\sim95$ $M_\odot$ yr$^{-1}$. The median SFR of the entire
sample is $\sim8$ $M_\odot$ yr$^{-1}$.  Although we assumed here that
the LBGs are dust-free, one could multiply the SFRs by a factor of 2.5
to correct for an average LBG extinction of $E(B-V)\approx0.1$ mag
(see next section) giving a median SFR of $\sim20$ $M_\odot$
yr$^{-1}$.

\subsubsection{UV Continuum colors}
\label{sec:uvslope}

\begin{figure}[t]
\begin{center}
\includegraphics[width=\columnwidth]{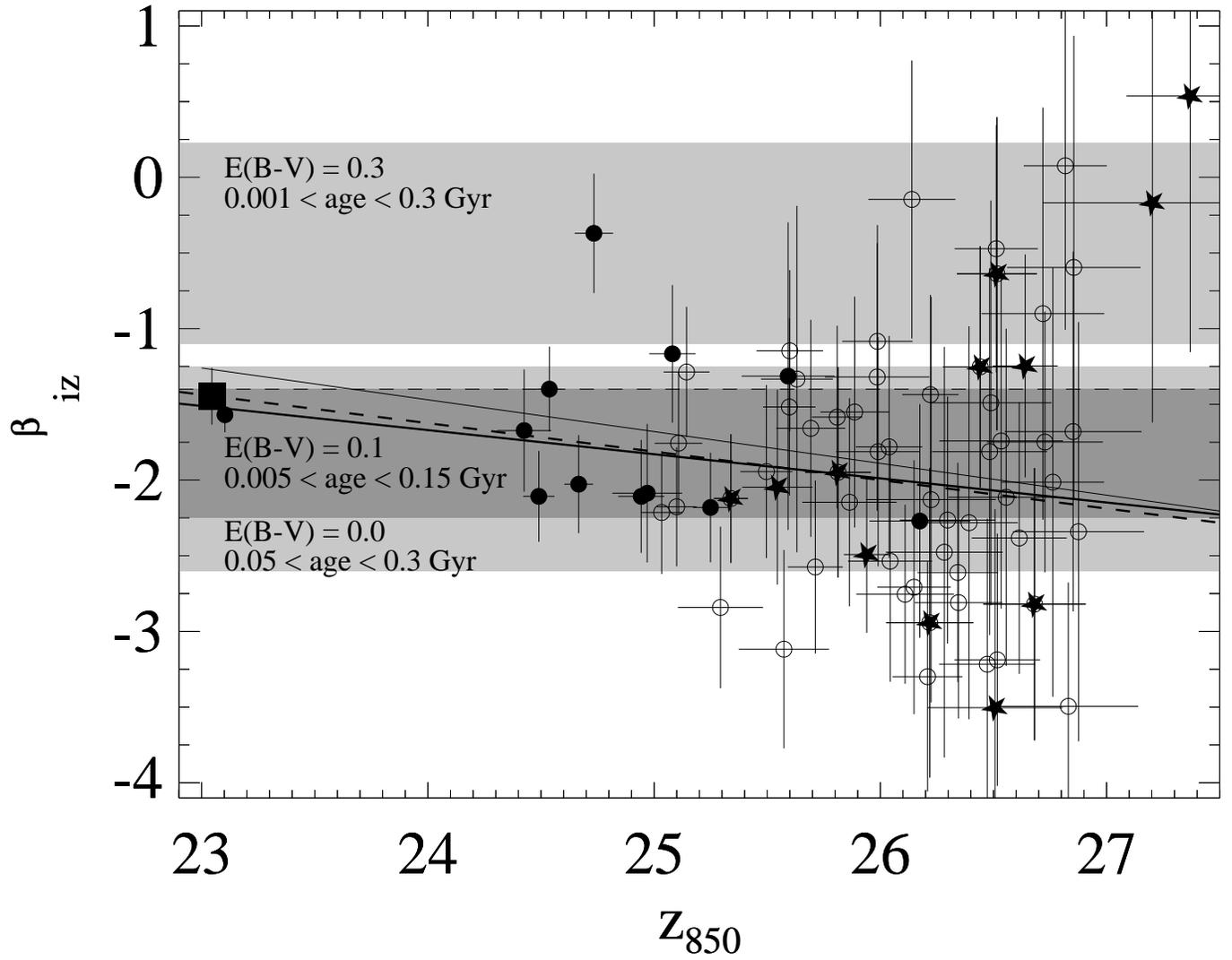}
\end{center}
\caption{\label{fig:lbgsbetas}\zp\ versus $\beta_{iz}$ for
  \gp-dropouts detected/undetected in $K_S$ (filled/open circles),
  LAEs (stars), and the radio galaxy (square).  The best-fit linear
  relations are indicated (thick lines, see text for details). The
  thin solid line is the relation for \bp-dropouts (R.J. Bouwens,
  private communication). The best-fit SED from \citet{papovich01}
  redshifted to $z=4$ has $\beta_{iz}\approx-1.4$ (thin dashed line).
  Shaded regions are for $E(B-V)=0.0$ mag with ages between 0.05 and 0.3
  Gyr (bottom light shaded region), $E(B-V)=0.1$ mag with ages between
  0.005 and 0.15 Gyr (dark shaded region), and $E(B-V)=0.3$ mag with ages
  between 0.001 and 0.3 Gyr (top light shaded region), assuming an
  exponential star formation history ($\tau=10$ Myr) with $0.2Z_\odot$
  metallicity and a Salpeter IMF.}
\end{figure}

\noindent
We calculate the UV continuum slopes from the \ip--\zp\ color. This
color spans the rest-frame wavelength range from $\sim1400$ \AA\ to
$\sim2000$ \AA.  We assume a standard power-law spectrum with slope
$\beta$ ($f_\lambda\propto\lambda^\beta$, so that a spectrum that is
flat in $f_\nu$ has $\beta=-2$). We calculate
\begin{equation}
\beta_{iz}=\frac{\mathtt{log_{10}} \frac{Q_{850}}{Q_{775}} - 0.4 (i_{775}-z_{850})}{\mathtt{log_{10}} \frac{\lambda_{775}}{\lambda_{850}}} - 2~,
\end{equation}
where $\lambda_{775}$ and $\lambda_{850}$ are the effective bandpass
wavelengths, and $Q_{775}$ and $Q_{850}$ are the fractions of the
continuum fluxes remaining after applying the recipe for foreground
neutral hydrogen absorption of \citet{madau95}. The break at
rest-frame 1216 \AA\ only starts to enter the \ip-band for galaxies at
$z\gtrsim4.7$. Thus $Q_{775}$ and $Q_{850}$ are unity and $\beta$ will
be relatively independent of redshift for $3.5\lesssim
z\lesssim4.5$. The measured slopes are plotted in
Fig. \ref{fig:lbgsbetas}. Excluding the two brightest sources, we find
$\langle\beta_{iz}\rangle=-1.95$. This is significantly bluer than
that found by \citet{papovich01}, although it is consistent at the
bright magnitude end where the comparison with $L^*$ galaxies is
appropriate (thin dashed line).

We have modeled the dependencies of the slope on age and dust using an
exponential star formation history ($\tau=10$ Myr) with $0.2Z_\odot$
metallicity and a Salpeter IMF. For a constant $E(B-V)\approx0.0$ the
range of slopes favours ages in the range 50--300 Myr.  Although the
incompleteness for faint, relatively red objects could be higher than
for blue objects \citep[e.g.][]{ouchi04_lf}, a high dust content
($E(B-V)\approx0.3$) is incompatible with the majority of the slopes
observed.  A linear fit to the data gives a slope-magnitude relation
of $\beta_{iz}=(-0.16\pm0.05)(z_{850}-25~\mathrm{mag})-1.84\pm0.11$
(thick solid line), which remains virtually unchanged when we exclude
the two brightest objects (thick dashed line).  Our relation is in
good agreement with that of \bp-dropouts in GOODS (R.J. Bouwens,
private communication).  The best-fit relation spans ages in the range
5--150 Myr for a constant $E(B-V)\approx0.1$. A similar
slope-magnitude relation is also observed in other works
\citep{meurer99,ouchi04_lf} and may imply a mass-extinction or a
mass-metallicity relation rather than a relation with
age. Interpreting the slope-magnitude relation as a mass-extinction
relation implies $E(B-V)\approx0.13$ mag at \zp$\approx23$ mag and
$E(B-V)\approx0.0$ mag at \zp$\approx27$ mag for a fixed age of 70
Myr.

\subsubsection{Rest-frame UV to optical colors}

\begin{figure}[t]
\begin{center}
\includegraphics[width=\columnwidth]{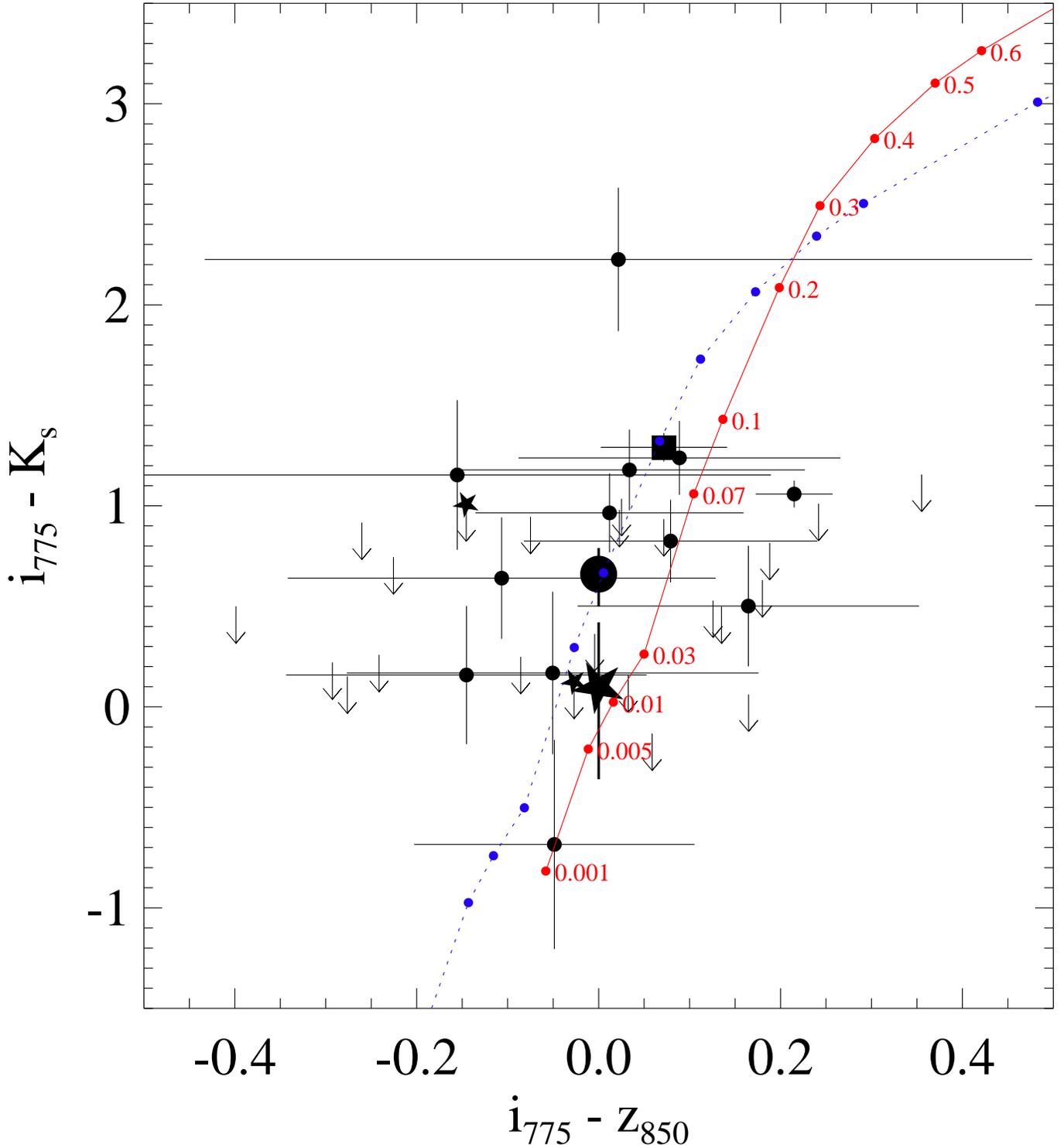}
\end{center}
\caption{\label{fig:uvopt1}Rest-frame UV-optical colors of the
  \gp-dropouts (circles), the LAEs L9 and L25 (stars), and the radio
  galaxy (square).  Arrows indicate $2\sigma$ limits for
  non-detections in $K_S$ (errors omitted for clarity).  Lines
  indicate the colors of a $\tau=10$ Myr SED ($0.2Z_\odot$) with ages
  in Gyr along the track for $E(B-V)=0.0$ mag (dotted) and $E(B-V)=0.15$ mag 
  (solid).  The large circle was obtained from a $K_S$-band stack of
  12 \gp-dropouts having 25.3$<$\ip$<$26.4 mag. The large star was
  obtained from a $K_S$-band stack of 5 LAEs within a similar
  magnitude range.  Their \ip--$K_S$ colors differ by $\sim0.7$ mag.}
\end{figure}

\noindent
At $z\sim4.1$, the filters \ip, \zp\ and $K_S$ probe the rest-frame at
$\sim$1500, 1800, and 4300\AA, respectively.  We detected 13 of the
\gp-dropouts in the $K_S$-band at $>2\sigma$. In Fig. \ref{fig:uvopt1}
we show the \ip--$K_S$ versus \ip--\zp\ color diagram. The \ip--$K_S$
color is more sensitive to the effects of age and dust than \ip--\zp,
due to its longer lever arm in wavelength. Comparing the colors to the
best-fit LBG spectrum from \citet{papovich01} redshifted to $z\sim4$
shows that the observed colors are consistent with ages in the range
10-100 Myr, although there will be degeneracy with
dust. Non-detections in the $K_S$-band suggests that more than 50\% of
the \gp-dropouts have ages less than 70 Myr, with a significant
fraction less than 30 Myr.  The radio galaxy is among the reddest
objects, although it has large gradients in \ip--$K_S$ among its
various stellar and AGN components \citep[see][]{zirm05}.

\begin{figure}[t]
\begin{center}
\includegraphics[width=\columnwidth]{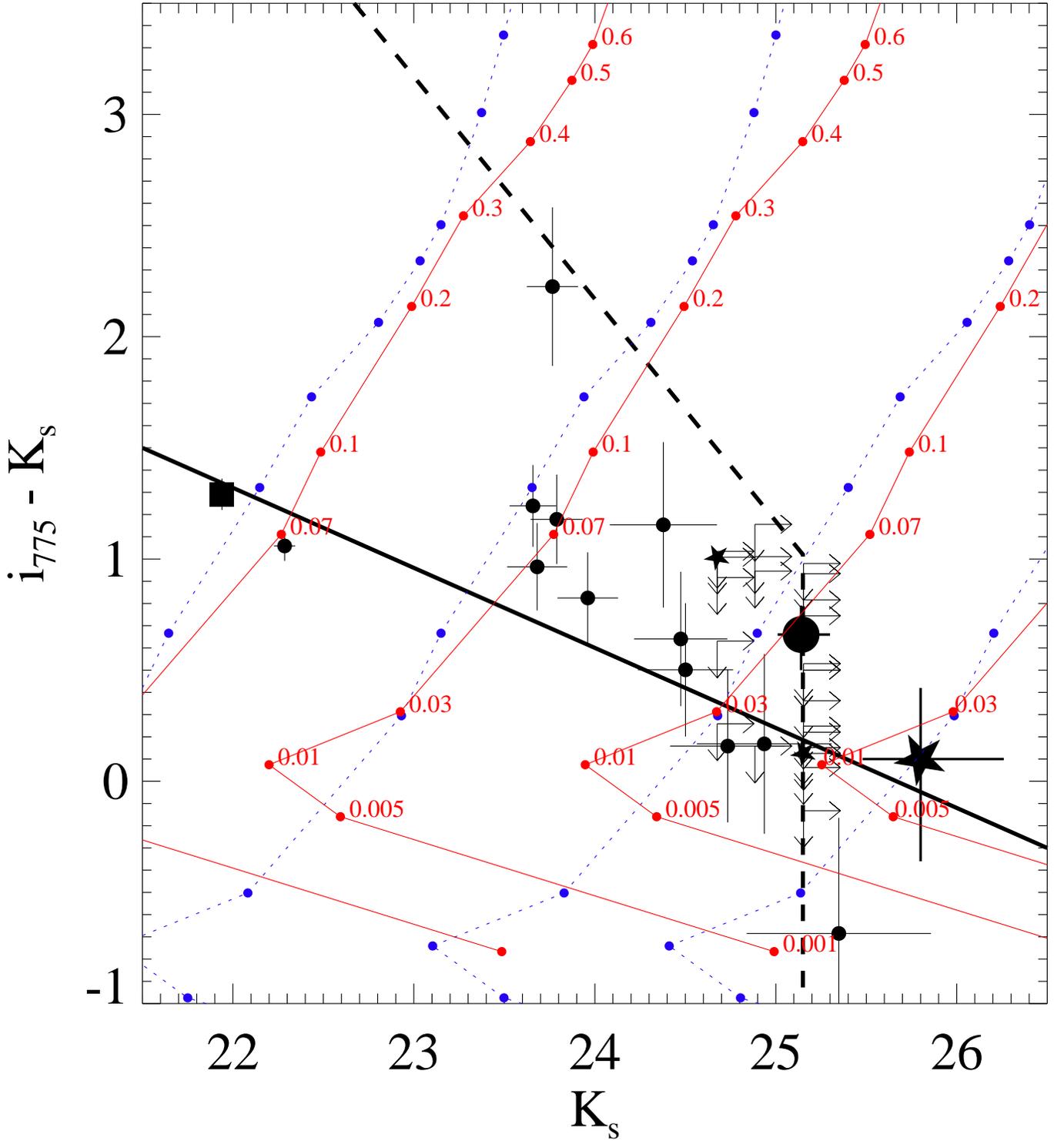}
\end{center}
\caption{\label{fig:uvopt2}Color-magnitude diagram of the \gp-dropouts
  (circles). The dashed line indicates the approximate $2\sigma$
  detection limits. The tracks are for $\tau=10$ Myr SEDs with
  different stellar masses of 0.03, 0.1, 0.5 and 2$\times10^{10}$
  M$_\odot$ for $E(B-V)=0.0$ mag (dotted) and $E(B-V)=0.15$ mag (solid). The
  thick solid line indicates the `blue envelope' of
  \citet{papovich04}, and suggests a color-magnitude relation in which
  luminosity correlates with either age or dust.  See the caption of
  Fig. \ref{fig:uvopt1} for further details.}
\end{figure}

In Fig. \ref{fig:uvopt2} we plot the \ip--$K_S$ versus $K_S$
color-magnitude diagram. \citet{papovich04} found evidence for a trend
of generally redder colors for galaxies that are brighter in $K_S$ in
GOODS. The effect is not likely to be a selection effect because the
objects are selected in the UV. \citet{papovich04} suggest that age or
dust of LBGs at $z\sim3-4$ may increase with increasing rest-frame
optical luminosity.  Our data are consistent with that conclusion.

A comparison with an exponentially declining starburst model track
from \citet{bruzualcharlot03} indicates that the stellar masses of the
objects in our sample span about one order of magnitude, ranging from
$\sim10^9$ M$_\odot$ for the faintest LBGs to $\sim10^{10}$ M$_\odot$
for the brightest.

\subsubsection{Sizes}

\noindent
We measured the half-light radius, $r_{hl}$, in \zp\ using SExtractor
by analysing the growth curve for each object out to $2.5\times r_\mathtt{Kron}$.  
Excluding the exceptionally large radio galaxy, the
measured radii range from unresolved ($\sim0\farcs07$) to $0\farcs42$,
corresponding to physical diameters of $\lesssim7$ kpc at
$z\sim4$. The average radius is $0\farcs17$ or $\sim1.4$ kpc. If we
divide our sample into two magnitude bins each containing an
approximately equal number of objects (achieved by placing a cut at
\zp$=$26.1 magnitude), the mean $r_{hl}$ are $0\farcs21\pm0\farcs01$
(error represents the standard deviation of the mean) and
$0\farcs14\pm0\farcs01$ in the bright and faint bins,
respectively. The difference is expected to be largely due to a larger
flux loss in the fainter sample (see Fig. \ref{fig:completeness}),
although fainter galaxies could indeed be smaller because of the
$r_{200}\sim V_c \sim L^{1/3}$ luminosity-size relationship, where
$r_{200}$ is the virial radius and $V_c$ is the circular dark matter
halo velocity \citep[see][]{mo98}.

\begin{figure*}[t]
\begin{center}
\includegraphics[width=\textwidth]{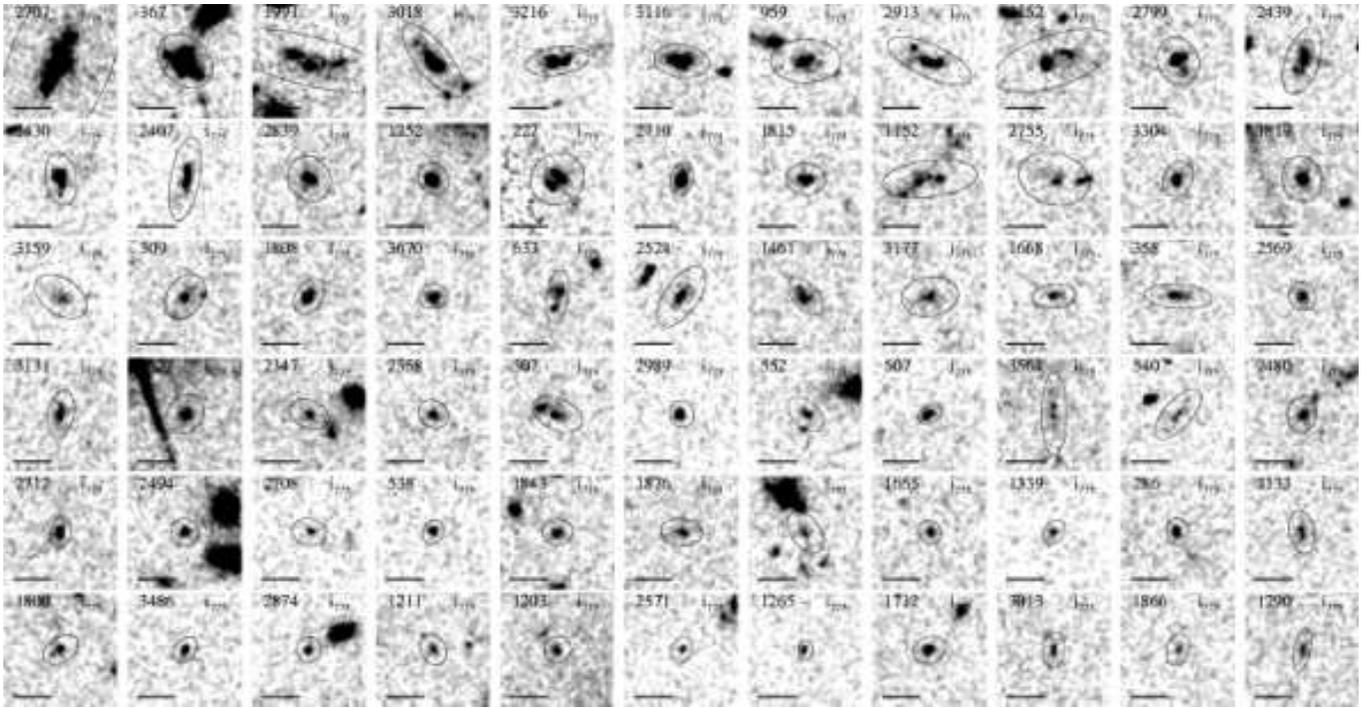}
\end{center}
\caption{\label{fig:lbgzstamps}\ip\ postage stamps
  ($3\arcsec\times3\arcsec$) of the \gp-dropout sample. The images
  have been smoothed using a Gaussian kernel of 0\farcs075
  (FWHM). Kron apertures are indicated. The scale bars measure
  1\arcsec.  North is up, East is to the left.}
\end{figure*}

The \ip-band morphologies of the \gp-dropouts are shown in
Fig. \ref{fig:lbgzstamps}. A separate section (Sect. \ref{sec:morphs})
is devoted to a non-parametric morphological analysis of our sample
and a comparable field sample.

\subsubsection{Point sources}

\noindent
The requirement for objects to be non-stellar to make it into our
sample had no big effect on the selection, since only 4 sources that
passed our color-color selection criteria were rejected based on high
stellarity index.  However, one of the point sources corresponds to a
spectroscopically confirmed \lya\ emitter (see next section),
indicating that at least one of the point sources is a genuine $z=4.1$
LBG. We derive an upper limit for the fraction of $z\sim4$ sources
missed that are unresolved (intrinsically small or AGN) of $\sim$6\%.

\subsection{\lya\ galaxies}
\label{sec:4.2}

\noindent
\citet{venemans02,venemans07} found an overdensity of 37 LAEs
($EW_{0,Ly\alpha}>15$\AA), all spectroscopically confirmed to lie
within $625\pm150$ km s$^{-1}$ of the radio galaxy $z=4.11$. All of
the 12 LAEs in the ACS field have been detected in \rp, \ip\ and \zp\
(see Table \ref{tab:laes}). Their properties are described below
\citep[see also][]{miley04}.

\subsubsection{Star formation rates}

\noindent
The \zp\ magnitudes are in the range 25.3--27.4 mag, corresponding to
a luminosity range of $\sim0.2-1.0L_*$.  The SFRs are $\sim1-14$
$M_\odot$ yr$^{-1}$ with a median of $5.1$ $M_\odot$ yr$^{-1}$ (not
including the effect of dust).  \citet{venemans05} calculated the SFRs
from \lya\ using SFR (M$_\odot$ yr$^{-1}$) $=8.7L_{Ly\alpha}$ (erg
s$^{-1}$)$/1.12\times10^{41}$, from \citet{kennicutt98} with the
standard assumption of case B recombination
\citep[][$L_{H\alpha}/L_{Ly\alpha}=8.7$ for gas that is optically
thick to \hi\ resonance scattering and no dust]{brocklehurst71}. In
general, we find good agreement between the SFRs calculated from the
UV compared to \lya\ with a median UV-to-\lya\ SFR ratio of 1.3.

\subsubsection{UV continuum colors}

\noindent
The UV slopes are indicated in Fig. \ref{fig:lbgsbetas} (stars).  The
slope can be constrained relatively well for the four brightest
emitters, which have $-2.1\pm0.4$, $-2.0\pm0.6$, $-1.9\pm0.7$, and
$-2.5\pm0.5$. The LAE slopes scatter around the $\beta_{iz}$-magnitude
relation for the \gp-dropouts found in Sect. \ref{sec:uvslope}, with a
sample average of $-1.7\pm1.2$.  These slopes are consistent with a
flat (in $f_\nu$) continuum, thereby favouring relatively low ages and
little dust.

\subsubsection{Rest-frame UV to optical colors}

\noindent
None of the 11 LAEs covered by the $K_S$ image were detected at the
$>2\sigma$ level.  We created a stack of the $K_S$-band fluxes for the
5 LAEs that lie in the deepest part of our NIR image. The subsample
had $25.3<$\zp$<26.4$ mag and
$\langle($\ip--\zp$)\rangle\approx0.0$. We obtained a $3\sigma$
detection for the stack finding $K_S=25.8^{+0.46}_{-0.32}$ mag and
hence \ip--\ks$\approx0.0$.  We compared this to 100 stacks of 5
\gp-dropouts each, selected from a sample of 12 that had a similar
range in \zp-band magnitudes and \ip--\zp\ colors as the LAEs. The
average detection among the 100 stacks was $\sim$5$\sigma$,
corresponding to $K_S=25.13^{+0.23}_{-0.18}$ mag and
\ip--\ks$\approx0.7$.  The results from the stacks have been indicated
in Figs. \ref{fig:uvopt1} and \ref{fig:uvopt2}. Although the
difference in the \ip--\ks\ color is significant at only
$\sim2\sigma$, one interpretation is that the faintest LAEs have
slightly lower stellar masses (a few times $10^8$ $M_\odot$) compared
to LBGs (see \se \ref{sec:discussion}).

\subsubsection{Sizes and morphologies}
\label{sec:lyasizes}

\noindent
If the \lya\ emission is associated with an extended halo of
outflowing gas, we might find differences in the typical radii of the
sources measured in the filter that includes \lya\ compared to pure
continuum filters.  We calculated $r_{hl,r}$ from the \rp-band, the
filter that includes \lya, and compared it to the $r_{hl,z}$ of the
continuum calculated from the \zp-band (Table \ref{tab:laes}).  The
mean $r_{hl}$ are $0\farcs13$ in \rp\ and $0\farcs12$ in \zp. At
$z=4.1$, the measured angular sizes correspond to physical radii of
$<3$ kpc, with a mean value of $\sim1$ kpc. We do not find evidence
for the sources to be more extended in \rp\ than they are in \zp,
suggesting that \lya\ emission either coincides with the continuum
region, or originates from a very extended, low surface brightness
halo not detected with ACS. One exception is source L7 which has
$r_{hl,r}=0\farcs18$ compared to $r_{hl,i}=0\farcs13$ and
$r_{hl,z}=0\farcs11$.

We have measured the $r_{hl}$ from a sample of 17 field stars in a
similar magnitude range.  The stars were selected on the basis of
SExtractor stellarity index of 1.0. Four of the LAEs (L4, L11, L20,
L22) have a $r_{hl}$ in both bands that is indistinguishable from that
of the stars. The UV luminosities of these unresolved LAEs are no
different than those of the resolved ones. Spectra show that the \lya\
lines are narrow ($<$1000 km s$^{-1}$) and no high ionization lines
have been detected \citep{venemans02}, ruling out that they are broad
emission line AGN. The light is probably due to unresolved stellar
regions with $r_{hl}\lesssim500$ pc. If we restrict ourselves to
resolved sources only, the mean $r_{hl}$ are $0\farcs15$ in both \rp\
and \zp.
 
\begin{figure}[t]
\begin{center}
\includegraphics[width=0.7\columnwidth]{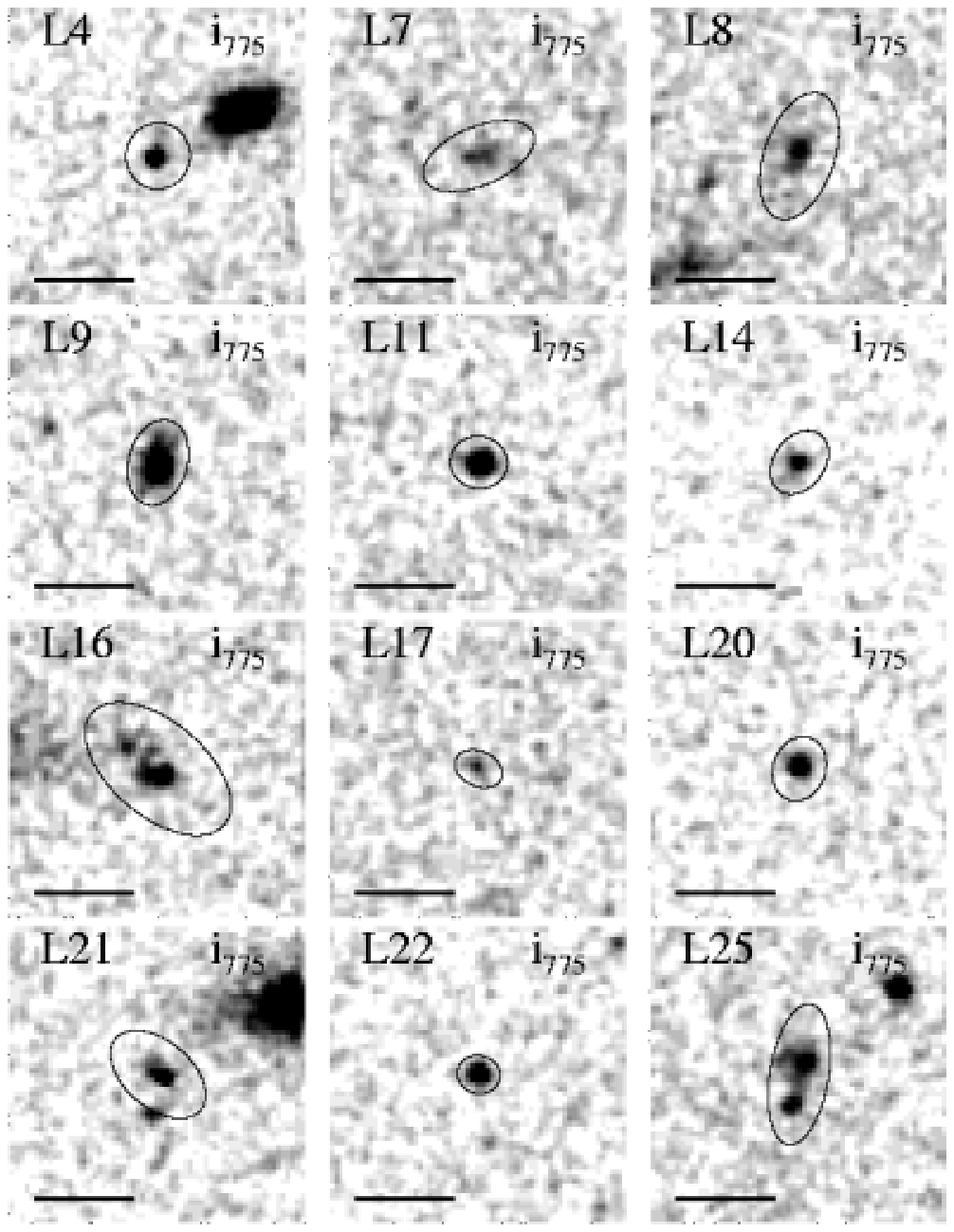}
\end{center}
\caption{\label{fig:lya}\ip\ postage stamps ($3\arcsec\times3\arcsec$)
  of the twelve spectroscopically confirmed \lya\ emitters of
  \citet{venemans02}. See the caption of Fig. \ref{fig:lbgzstamps} for
  details.}
\end{figure}

The ACS morphologies at rest-frame $\sim1500$\AA\ are shown in
Fig. \ref{fig:lya}.  Two sources (L16 \& L25) have double nuclei
separated by $\sim0\farcs5$ ($\sim3$ kpc) that are connected by faint,
diffuse emission suggestive of merging systems.

\section{Morphological analysis}
\label{sec:morphs}

\noindent
In order to quantify the wide range in ACS continuum morphologies seen
in Fig. \ref{fig:lbgzstamps}, we have carried out a nonparametric
morphological analysis of the \gp-dropout sample.  Following
\citet{lotz04} we determined the following morphological coefficients:
1) The Gini coefficient ($G$), a statistic for the relative
distribution of an object's flux over its associated pixels, 2)
$M_{20}$, the normalised second order moment of the brightest 20\% of
a galaxy's pixels, and 3) concentration ($C$), the ratio of the
circular radii containing 20\% and 80\% of the total flux
\citep[see][and references therein]{conselice03}.

To improve the $S/N$ of our sample we coadded the \ip\ and \zp\
images, giving a total exposure time of 23500 s.  To further improve
the $S/N$ per pixel the images were binned using a $2\times2$ binning
scheme.  Pixels were flagged as belonging to an object if they were
inside one `Petrosian radius' \citep{petrosian76}. This ensures that
the morphological analysis is relatively insensitive to varying
surface brightness limits and $S/N$ among different objects
\citep{lotz04}. The number of \gp-dropouts with sufficient
$(S/N)_{pixel}$ was maximized by setting the free Petrosian parameter
$\eta$ to 0.3. We measured the morphologies for a total of 15 of the
\gp-dropouts that had $(S/N)_{pixel}>3$, which included the radio
galaxy. The coefficients were all determined within a maximum radius
of $1.5\times r_p$.  We used SExtractor's segmentation maps to mask
out all pixels suspected of belonging to unrelated sources. Errors on
the coefficients were determined using Monte Carlo simulations. The
value of each pixel was modified in such a way that the distribution
of values were normally distributed with a standard deviation equal to
that given by the RMS image value for the corresponding pixel. We
remark that the concentration ($C$) is often an underestimate of the
true concentration of high redshift objects, because the 20\% flux
radius is often smaller than the $\sim0.1\arcsec$\ resolution obtained
by HST\footnote{We did not apply the sub-pixel rebinning technique
  used by \citet{lotz05} that attemps to partially correct for this
  effect.}.

\begin{figure}[t]
\begin{center}
\includegraphics[width=\columnwidth]{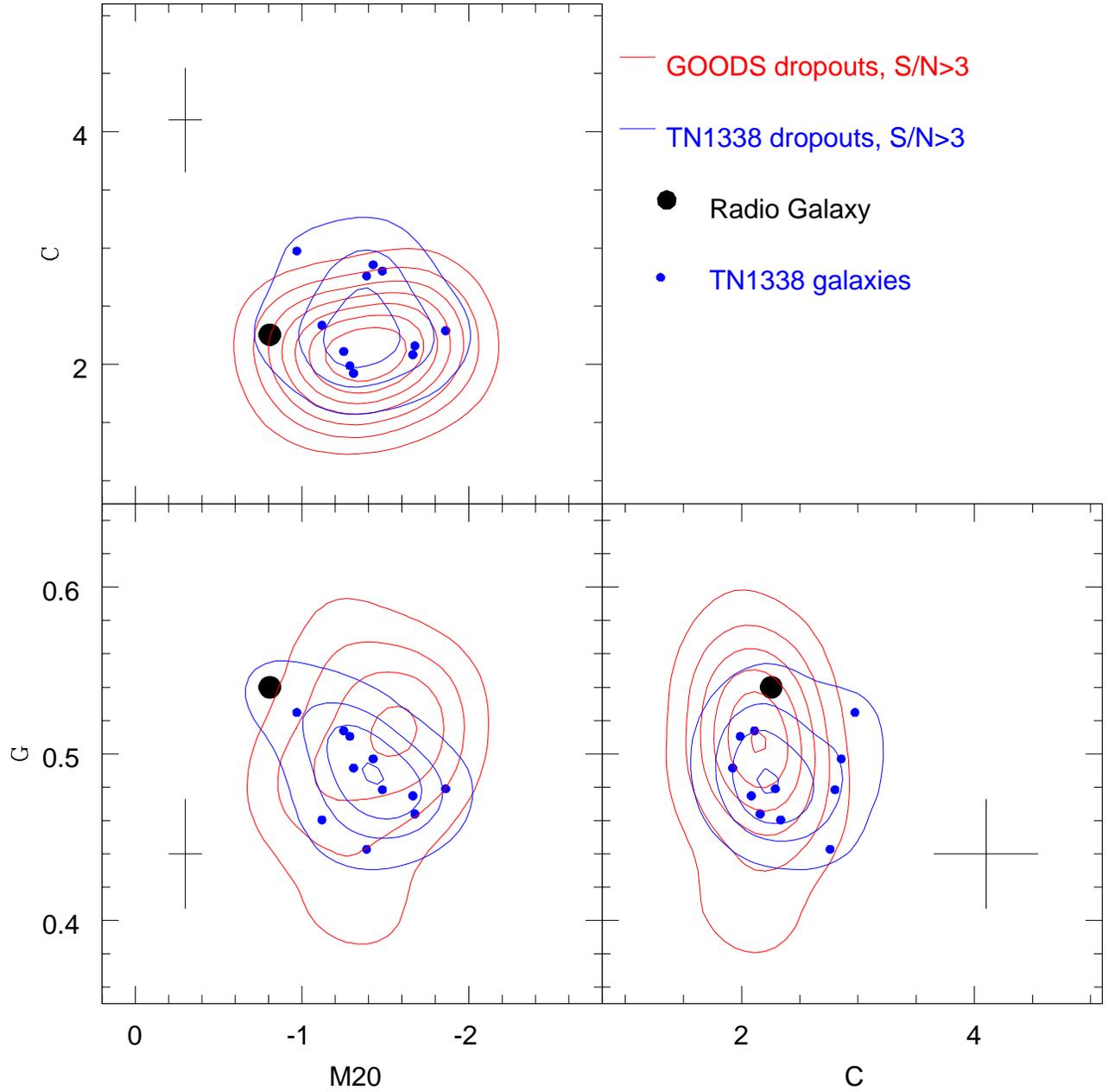}
\end{center}
\caption{\label{fig:nick}The morphological Gini coefficient ($G$),
  $M_{20}$, and concentration ($C$) for the \gp-dropouts in TN1338
  (blue points and contours). The parameter distributions determined
  from $\sim$70 LBGs at $z\sim4$ selected in the GOODS CDF-S field are
  shown for comparison (red contours). The radio galaxy is indicated
  by the large black circle.  The approximate size of the error bars
  on the measurements are indicated in a corner of each panel.}
\end{figure}

In Fig. \ref{fig:nick} we plot the parameters measured for the TN1338
\gp-dropouts (blue points). The radio galaxy (large circle) has
non-average values in each of the parameter spaces, owing to its
complex morphology as described in detail by \citet{zirm05}.  We have
applied a similar morphological analysis to a sample of 70 LBGs at
$z\sim4$ with $(S/N)_{pixel}>3$ selected from the CDF-S GOODS field
for comparison.  Fig. \ref{fig:nick} indicates that both the centroids
and the spread of the TN1338 parameter distributions (blue contours)
coincide with that of the parameter distributions determined from
GOODS (red contours).

\section{Evidence for an overdensity associated with TN J1338-1942 at $z=4.1$?}
\label{sec:overdense}

\subsection{Surface density distribution}

\begin{figure}[t]
\begin{center}
\includegraphics[width=0.7\columnwidth]{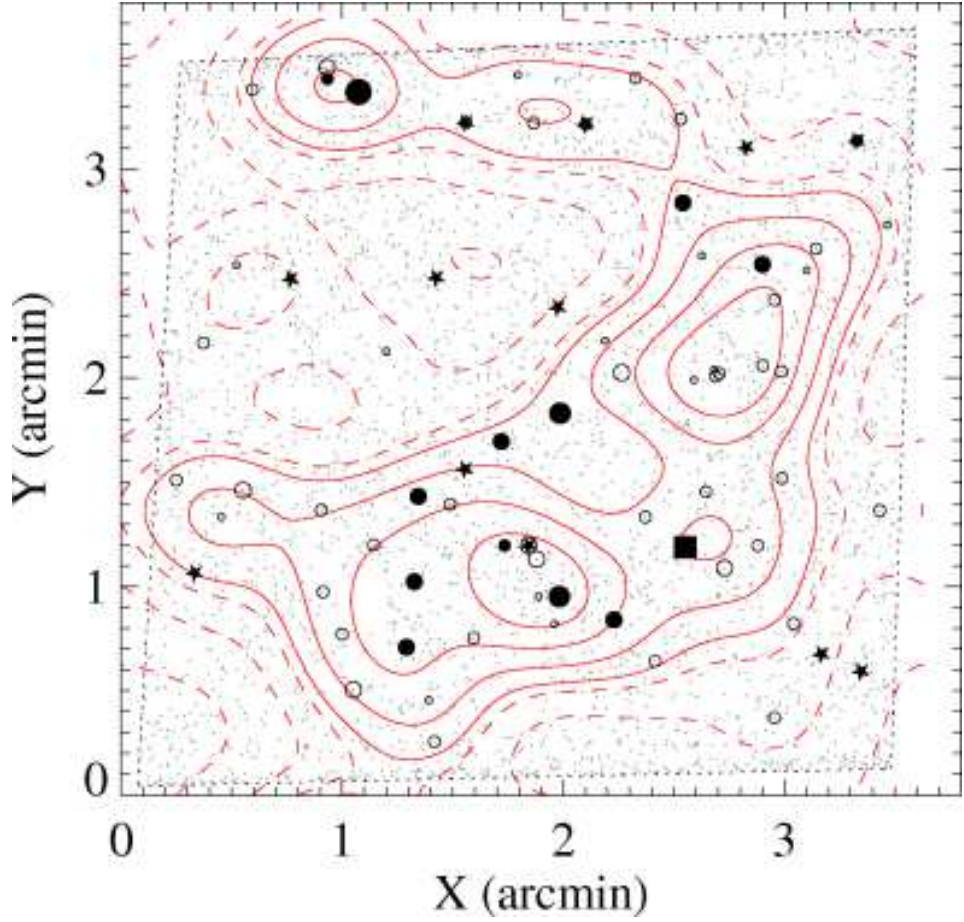}
\end{center}
\caption{\label{fig:surfdens}Object map of the \gp-dropout candidates
  (circles), TN J1338--1942 (square), \lya\ emitters (stars) and the
  detection catalog (points). \gp-dropouts detected in $K_S$ are
  indicated by filled circles. Larger circles indicate brighter
  objects in \zp. The contours represent density fluctuations
  $\Delta\equiv(\Sigma-\bar{\Sigma})/\bar{\Sigma}$ of --1, --0.5,
  --0.1 (dashed contours) and +0.1, +0.5, +1, +1.5 (solid contours),
  achieved by smoothing the object map with a Gaussian of width
  36\arcsec, or 250 kpc (FWHM), using equal weights. LAEs that are not
  in the \gp-dropout sample were not included in the density
  contours.}
\end{figure}

\noindent
In Fig. \ref{fig:surfdens} we show the angular distribution of the
\gp-dropouts in TN1338. Contours of the local object surface density
compared to the average field surface density illustrate that the
\gp-dropouts lie predominantly in a filamentary structure. Another
density enhancement is located near the top edge of the
image. Interestingly, both peaks in the object surface density
distribution coincide with an extremely bright $\sim 6L^*$ LBG. One of
these is the radio galaxy, which takes a central position in the
largest concentration of \gp-dropouts in the field. About half of the
dropouts lies in a $2\farcm1\times2\farcm1$ region that includes the
radio galaxy.

The \gp-dropouts detected in $K_S$ are found only in the high density
regions, most notably in the clump to the left of the radio
galaxy. The filamentary distribution of the dropouts is not seen in
the spatial distribution of the 12 LAEs, which are distributed more
uniformly over the field.  In fact, 8 of the emitters lie in regions
that are underdense compared to the overall distribution of
\gp-dropouts.

\subsection{Comparison with `field' LBGs from GOODS}
\label{sec:compgoods}

\noindent
Here we will test whether the structure of \gp-dropouts represents an
overdensity of star-forming galaxies associated with TN J1338--1942,
similar to the overdensity of LAEs discovered by
\citet{venemans02,venemans07}.  To determine the `field' surface
density of \gp-dropouts we have extracted a control sample by applying
our selection criteria to the simulated images based on \bp-dropouts
in the CDF-S GOODS field as described in Sect. 2.7.1, and we recall
from Sect. 2.7.2 that these image transformations have been shown to
be representative at the $\gtrsim80\%$ confidence level.

\begin{figure}[t]
\begin{center}
\includegraphics[width=\columnwidth]{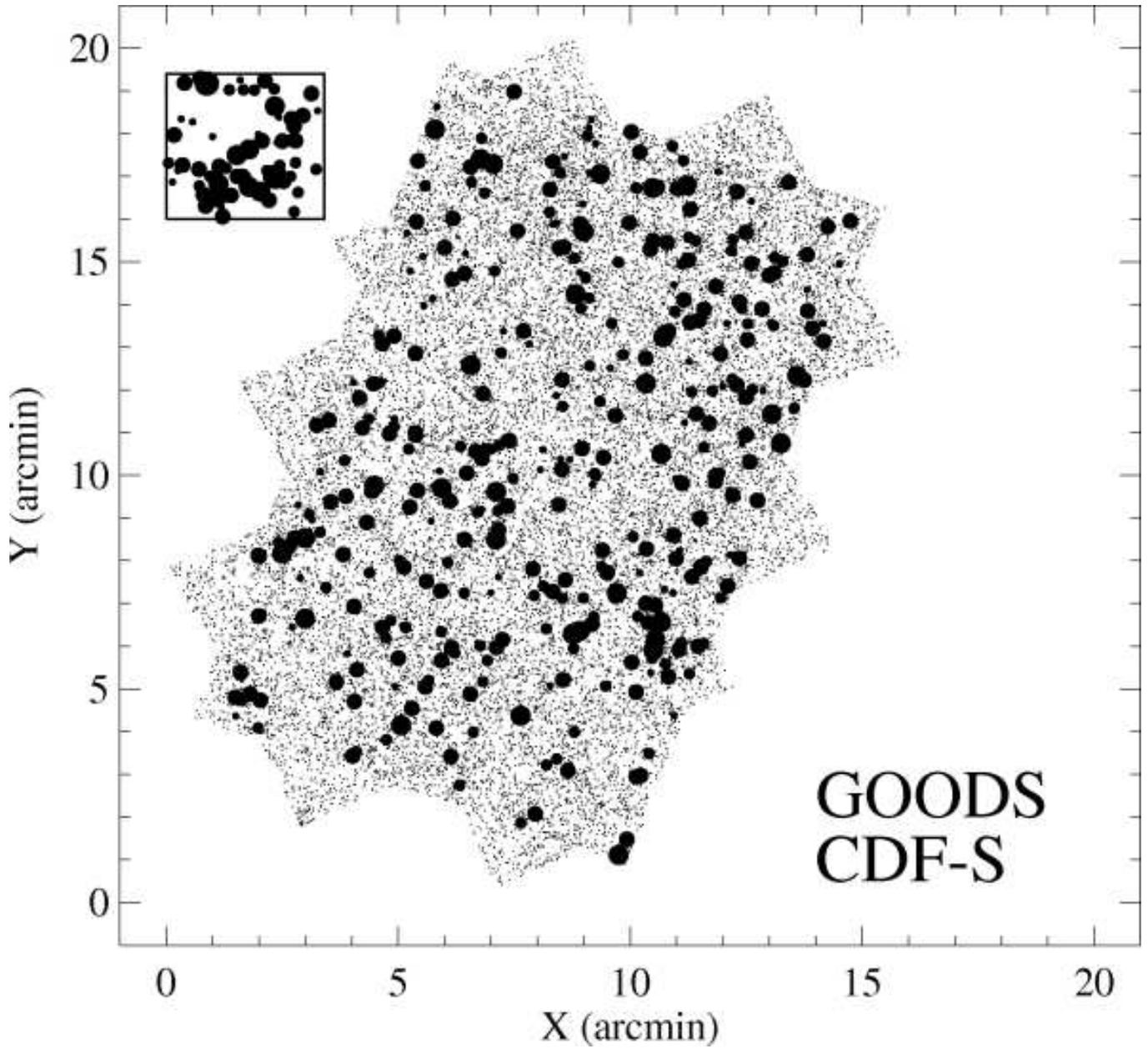}
\end{center}
\caption{\label{fig:mosaic}Distribution of $z\sim4$ LBGs in the CDF-S
  GOODS field with \zp$\le27.0$ mag (circles).  Larger symbols correspond
  to brighter objects. The inset in the top left shows the
  $3\farcm4\times3\farcm4$ TN1338 field and the distribution of
  \gp-dropouts at the same scale as the CDF-S GOODS field for
  comparison.}
\end{figure}

A mosaic of the CDF-S GOODS tiles is shown in Fig. \ref{fig:mosaic}.
At \zp$<27.0$ mag there are a total of 361 \gp-dropouts in the
transformed CDF-S in an area of 159 arcmin$^{2}$, giving an average
surface density of 2.27 arcmin$^{-2}$ (see Fig. \ref{fig:mosaic}). The
surface densities are 1.82 arcmin$^{-2}$ and 1.16 arcmin$^{-2}$ for
\zp$<26.5$ and \zp$<26.0$ mag, resp.  The surface density of
\gp-dropouts in TN1338 is approximately $2.5\times$ higher for each
magnitude cut (5.64 arcmin$^{-2}$, 4.36 arcmin$^{-2}$, and 2.74
arcmin$^{-2}$, resp.).

\begin{figure}[t]
\begin{center}
\includegraphics[width=0.7\columnwidth]{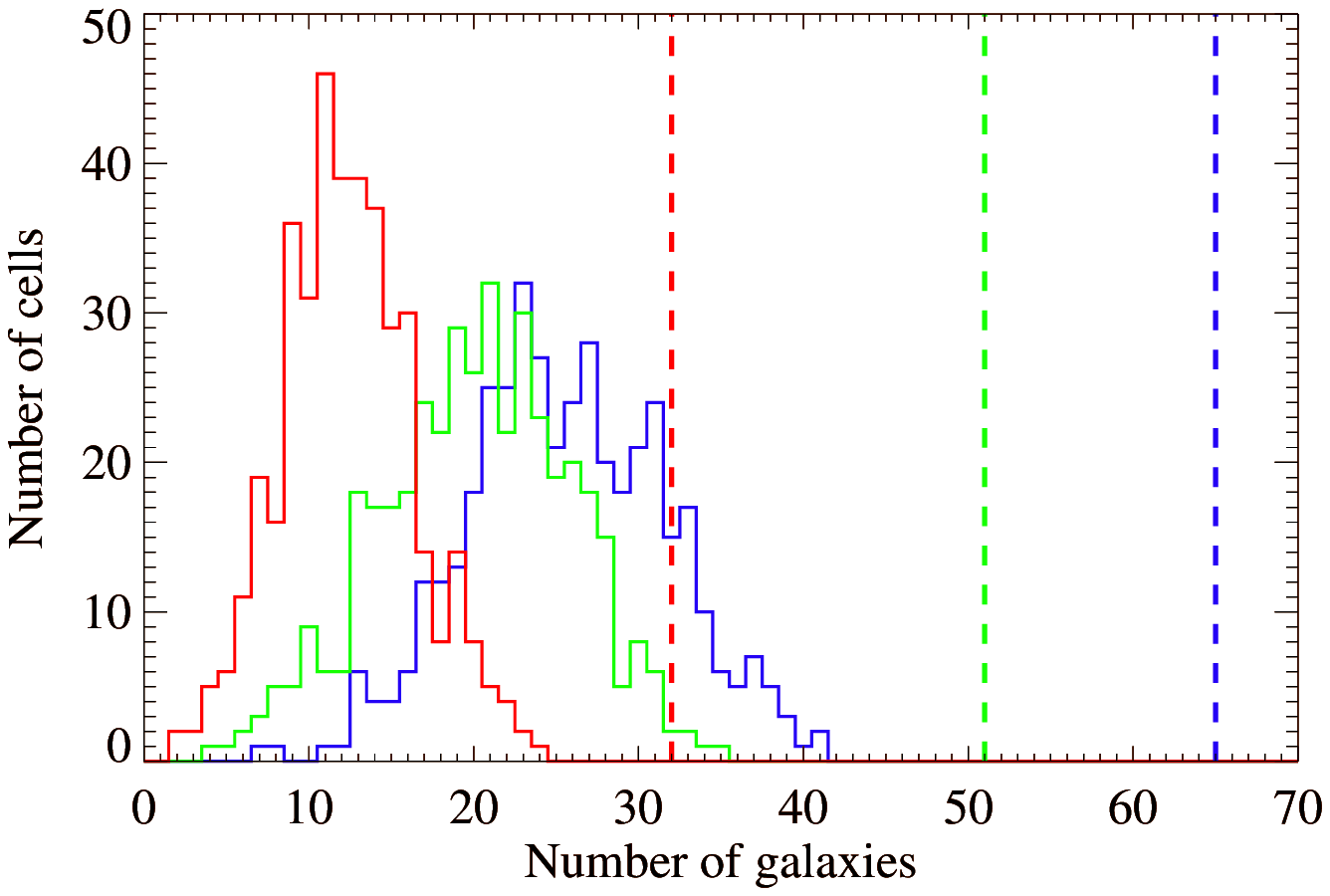}\\
\includegraphics[width=0.7\columnwidth]{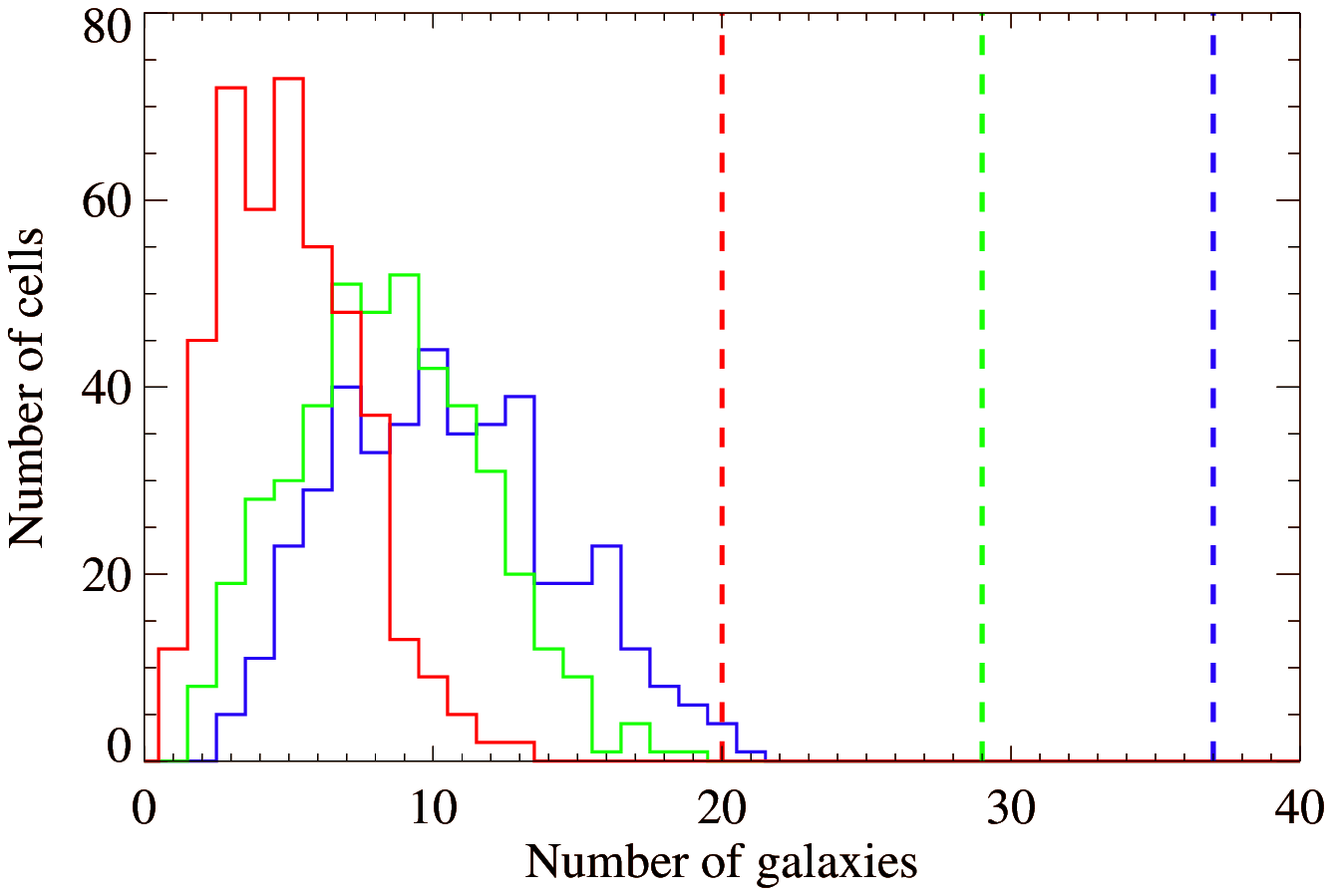}
\end{center}
\caption{\label{fig:cells}Counts in cells analysis of $z\sim4$ LBGs in
  the GOODS simulations compared to TN1338. {\it Top panel:}
  Histograms of the number of objects in square cells the size of
  TN1338 ($3\farcm4\times3\farcm4$) for \zp$<$27.0 (blue), 26.5
  (green) and 26.0 mag (red). The number of \gp-dropouts in TN1338 are
  indicated by the vertical lines of corresponding color.  {\it Bottom
    panel:} Same as top, but for $2\farcm1\times2\farcm1$ cells. The
  number of \gp-dropouts in TN1338 exceeds the number encountered in
  GOODS for each limiting magnitude and cell size.}
\end{figure}

What is the significance of this factor 2.5 surface overdensity?  LBGs
belong to a galaxy population that is strongly clustered at every
redshift \citep{porciani02,ouchi04_r0,lee05}, with non-negligible
field-to-field variations. In our particular case, it is interesting
to estimate the chance of finding a particular number of \gp-dropouts
in a single $3\farcm4\times3\farcm4$ ACS pointing. Analysing each of
the 15 GOODS tiles individually, the lowest number of \gp-dropouts
encountered was 12, and the highest was 37 to \zp=27.0 mag. Next, we
measured the number of objects in $\sim500$ randomly placed, square 11
arcmin$^{2}$ cells in the CDF-S GOODS mosaic. The cells were allowed
to overlap so that the chance of finding the richest pointing possible
was 100\%.  In Fig. \ref{fig:cells} ({\it top panel}) we show the
histogram of counts-in-cells for the three different magnitude
cuts. In each case the number of objects in TN1338 (indicated by the
dashed lines) falls well beyond the high-end tail of the distribution,
with none of the cells randomly drawn from GOODS containing as many
objects (the highest being 41, 35, and 24 for \zp$<$27.0, 26.5, 26.0
mag). Approximating the distributions with a Gaussian function
(strictly speaking, this is only valid in the absence of higher order
clustering moments, as well as non-linear clustering at very small
scales), we find a surface overdensity of 2.5 at $5-6\sigma$
significance with respect to the simulated CDF-S. The measured
standard deviations were corrected by 4\% to take into account that
the counts in cells distribution will appear narrower due to the fact
that our control field is not infinitely large.

Comparable significance for an overdensity is found if we focus on a
smaller region of 4.4 arcmin$^2$, where more than half of the
\gp-dropouts are located (see Fig. \ref{fig:surfdens}). Drawing
$2\farcm1\times2\farcm1$ regions from the CDF-S GOODS field
(Fig. \ref{fig:cells}, {\it bottom panel}) yielded a maximum of 21, 19
and 13 objects for the three magnitude cuts, resp. The region in
TN1338 corresponds to surface overdensities of 3.4, 3.2, and 4.0, but
with $\sim5\sigma$ significance due to the fact that the
counts-in-cells distribution function is much wider due to the
relatively small cell size compared to the fluctuations in the surface
density of LBGs (bottom panel).

We have not applied the counts in cells to simulations of the GOODS
HDF-N, since it has been shown that the northern GOODS field is
$\sim10$\% less rich in \bp-dropouts compared to its southern
counterpart \citep{bouwens07}. We conclude that the number of
\gp-dropouts in TN1338 represents a highly significant overdensity
with respect to the 314 arcmin$^2$ GOODS fields. Below we will further
investigate implications of large-scale structure and cosmic variance.

\subsection{\wt\ and sub-halo clustering}

\noindent
The two-point angular correlation function (ACF) is one of the most
powerful tools to study the large-scale distribution of high redshift
galaxies.  Recent measurements of the clustering of LBGs at $3<z<5$
show that the ACF, \wt, deviates from the classical power-law at small
angular scales. This behaviour is expected in the regime where
non-linear clustering within single halos dominates over the
large-scale clustering between halos.  This effect has now been shown
to be present in large LBG samples \citep{ouchi05c,lee05}.

\begin{figure}[t]
\begin{center}
\includegraphics[width=\columnwidth]{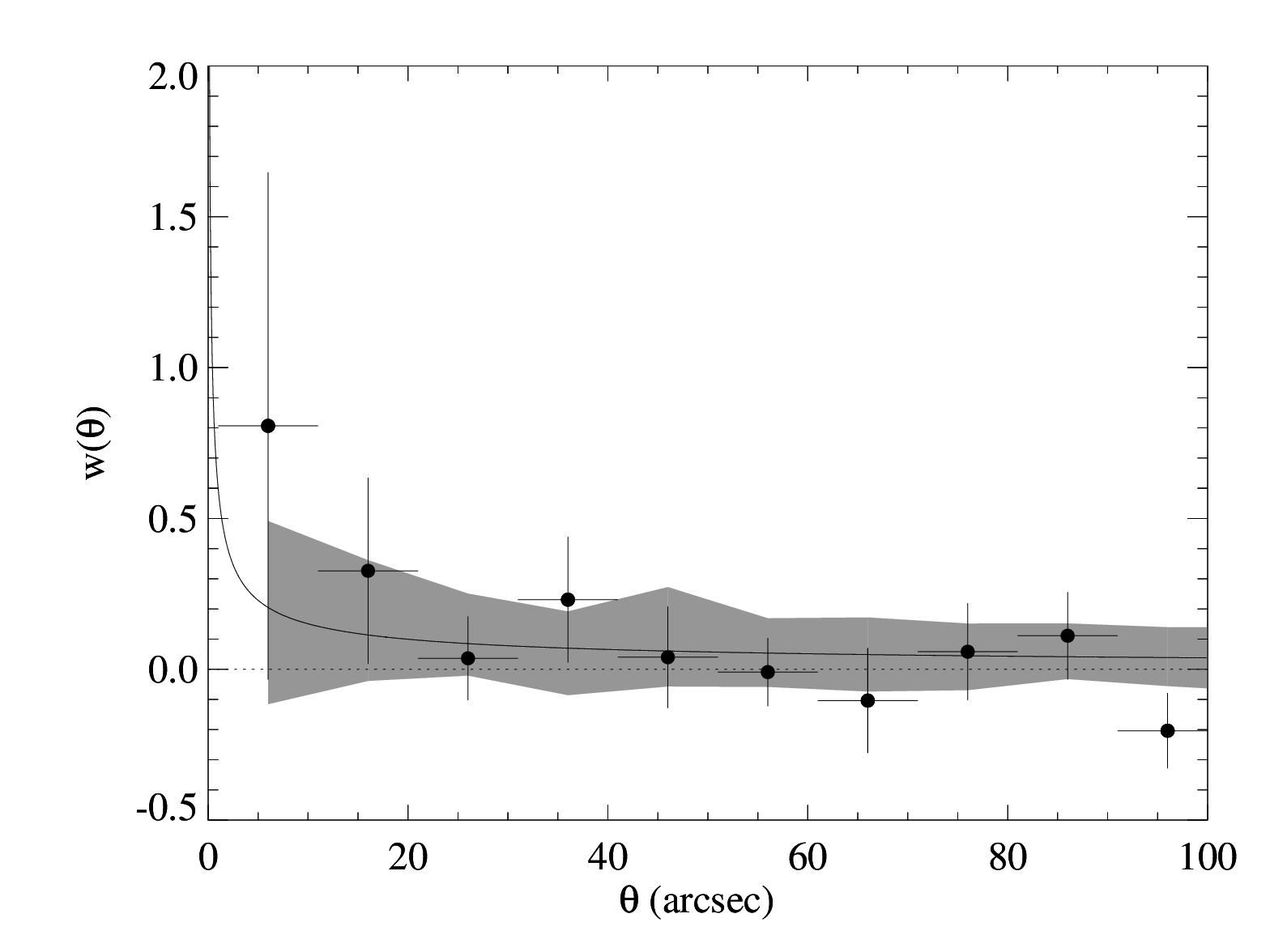}
\end{center}
\caption{\label{fig:wt}The angular two-point correlation function for
  \gp-dropouts in the TN1338 field (solid points).  The solid line
  indicates the angular correlation function
  $w(\theta)=0.6\theta^{-0.6}$, as measured for \bp-dropouts with
  \zp$\lesssim26.5$ mag in GOODS \citep{lee05}. The shaded region
  indicates the $1\sigma$ spread in \wt\ at each $\theta$ that was
  measured among 25 $3\farcm4\times3\farcm4$ fields of the same
  geometry and source density as the TN1338 field extracted from a
  $17\arcmin\times17\arcmin$\ mock catalog with
  $w(\theta)\approx0.6\theta^{-0.6}$.}
\end{figure}

The structure found in TN1338 provides a unique opportunity to test
the contribution of the one-halo term to \wt\ within a single,
overdense field. We measured \wt\ using the estimator 
$w(\theta)=[A_1DD(\theta)-2A_2 DR(\theta)+RR(\theta)]/RR(\theta)$ with
$A_1=N_r(N_r-1)/(N_g(N_g-1))$ and $A_2=(N_r-1)/2N_g$
\citep{landy93}. We used 25 random catalogs of 10000 sources each, and
the errors on \wt\ were estimated from the standard deviation among 32
bootstrap samples of the original data \citep{ling86}.  When the field
size is relatively small, the average density of a clustered
distribution is overestimated because the field where the clustering
is being measured is also the field where the average density has to
be estimated from.  We therefore estimated the integral constraint
(IC), $IC/A_w=\sum_iRR(\theta_i)\theta_i^{-\beta}/\sum_iRR(\theta_i)=0.073$,
where we assumed a fixed slope of $\beta=0.6$.  The result is plotted
in Fig. \ref{fig:wt}. We used bins of 10\arcsec, but excluded the
separations of $\theta<1\arcsec$.  Since we do not expect any signal
in the large-scale, 2-halo clustering due to the limited size of our
sample, we have safely applied the IC to the datapoints using the
large-scale clustering amplitude $A_w\approx0.6$ for B-dropouts in
GOODS at \zp$\lesssim26.5$ mag measured by \citet{lee05}. \wt\ is
consistent with no clustering given the large bootstrap errors. The
datapoint at $\theta<10\arcsec$ lies above the expected large-scale
clustering amplitude (solid line). The scale agrees well with the
expected location of the upturn in \wt\ due to sub-halo clustering
\citep{ouchi05c,lee05}.

To test the significance of possible sub-halo clustering in the field
of TN1338, we constructed a mock field having large-scale clustering
properties resembling those of LBGs at $z\sim4$. We used the formalism
of \citet{soneira78} to create an object distribution with a choice
two-point ACF. The procedure is as follows. First a random position is
chosen. This forms the center of a pair of points that are placed with
a random position angle and separation $\theta_1$. Each point forms
the center for a new pair with separation $\theta_2=\theta_1/\lambda$
and random position angles. This process is repeated until $L$ levels,
each level contributing $2^L$ points with separations
$\theta_1/\lambda^{L-1}$ to the `cluster'.  Next, a new cluster center
is randomly chosen in the field, and the cluster is again populated
with a depth of $L$ levels.  This is repeated until the mock field
contains $N$ clusters. The resulting point distribution will have a
power-law two-point ACF with its slope determined by the choice of
$\lambda$, and its smallest and largest angular scales determined by
the point separations at the first and the last levels,
respectively. Because we need both many levels to get sufficient
signal in \wt\ at all angular scales, and many clusters to get
sufficient coverage of the area, the method above produces far too
many points at first. It is therefore common to introduce a parameter,
$f$, which is the probability that a point makes it into the final
sample when drawing a random subsample.  We then calculate the number
of clusters $N$ necessary to match a particular surface density given
this $f$.  The amplitude of the ACF, $A_w$, solely depends on the
choice of $f$, since the clusters are randomly distributed with
respect to each other. We iteratively created mock samples with
different $f$ and measured \wt\ until the best-fit amplitude matched
the amplitude of the ACF that we wish to model (corrected for the
IC). The size of the mock field was set to $17\arcmin\times17\arcmin$
with a surface density of $\approx5$ arcmin$^{-2}$ to match the
density found in TN1338.  To model the result of \citet{lee05},
$w(\theta)\approx0.6\theta^{-0.6}$, we required an $f$ of 0.001 (with
$\theta_1=8\farcm3$ and $L=13$).  Having modeled the observed
two-point statistics successfully, we extracted 25
$3\farcm4\times3\farcm4$ `ACS' fields from the mock sample and
measured the mean $w(\theta)$ and its standard deviation. The result
is indicated in Fig. \ref{fig:wt} (shaded region).  The mean \wt\
corresponds to the large-scale clustering that was built into the
larger mock field (solid line).  For $\theta<10\arcsec$, the error
determined from the mock simulations is about half the bootstrap
error, and there is a $2\sigma$ discrepancy between the clustering
observed in TN1338 and the expected clustering of a similarly sized
mock field. We interpret this excess at small scales in the TN1338
field as likely being due to sub-clustering of galaxies that are
physically interacting on scales comparable or smaller than their
typical halo sizes (see Sect. \ref{sec:halosizes}).

\subsection{Spectroscopy and photometric redshifts}

\noindent
Excluding the radio galaxy, 6 of the LAEs confirmed by
\citet{venemans02,venemans07} are also in our photometrically selected
LBG candidate sample. These large equivalent width \lya\ LBGs lie in a
narrow redshift interval ($\Delta z\approx0.03$) centered on the
redshift of the radio galaxy. We further obtained spectroscopic
redshifts for three of the candidate LBGs in our sample.  The spectrum
of the $\sim6L^*$ object \#367 shows several absorption lines typical
for LBGs at a redshift of $3.830\pm0.002$
(Fig. \ref{fig:spectrum}). Object \#3018 was found to have a redshift
of $z=3.911\pm0.004$, based on the presence of (faint) \lya\ in
emission, and \oisiii $\lambda$1303 and \cii $\lambda$1334 in
absorption. Candidate \#959 has strong \lya\ (as confirmed by its
asymmetry) at a redshift of $z=3.92\pm0.01$. Although the redshifts of
these three LBGs in particular indicate no physical association with
the radio galaxy and \lya\ emitters, the spectroscopic results confirm
that our \gp-dropout selection criteria successfully identify LBGs at
$z\approx4$.

\begin{figure}[t]
\begin{center}
\includegraphics[width=0.7\columnwidth]{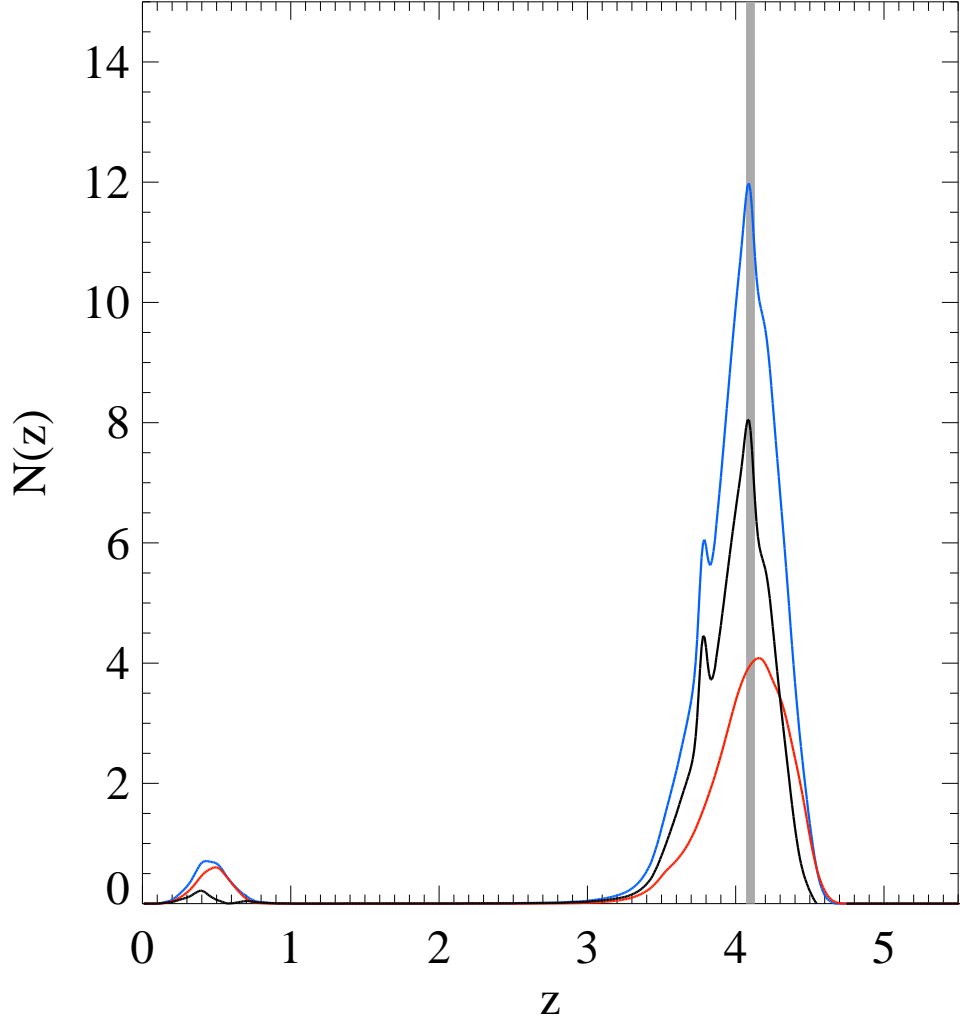}
\end{center}
\caption{\label{fig:bpzprobs}Total $z_B$ probability distributions for
  \gp-dropouts in TN1338 (blue) and in the CDF-S GOODS field
  (red). The area under the blue curve is equal to the number of
  objects in the TN1338 sample. The red curve for GOODS has been
  normalised to the area of the TN1338 field. The GOODS curve was
  subtracted from the TN1338 curve to bring out the residuals of the
  redshift distribution for TN1338 (black curve). The residual 
  peaks at the redshift of the radio galaxy and LAEs (shaded
  region). A secondary peak of 1 object lies at $z_B\approx3.8$, the
  redshift of the other $\sim6L^*$ object in the field. The peaks at
  $z\sim0.5$ are due to the alternative probability that a fraction of
  the \gp-dropouts could be 4000\AA\ break objects.}
\end{figure}

\begin{figure}[t]
\begin{center}
\includegraphics[width=\columnwidth]{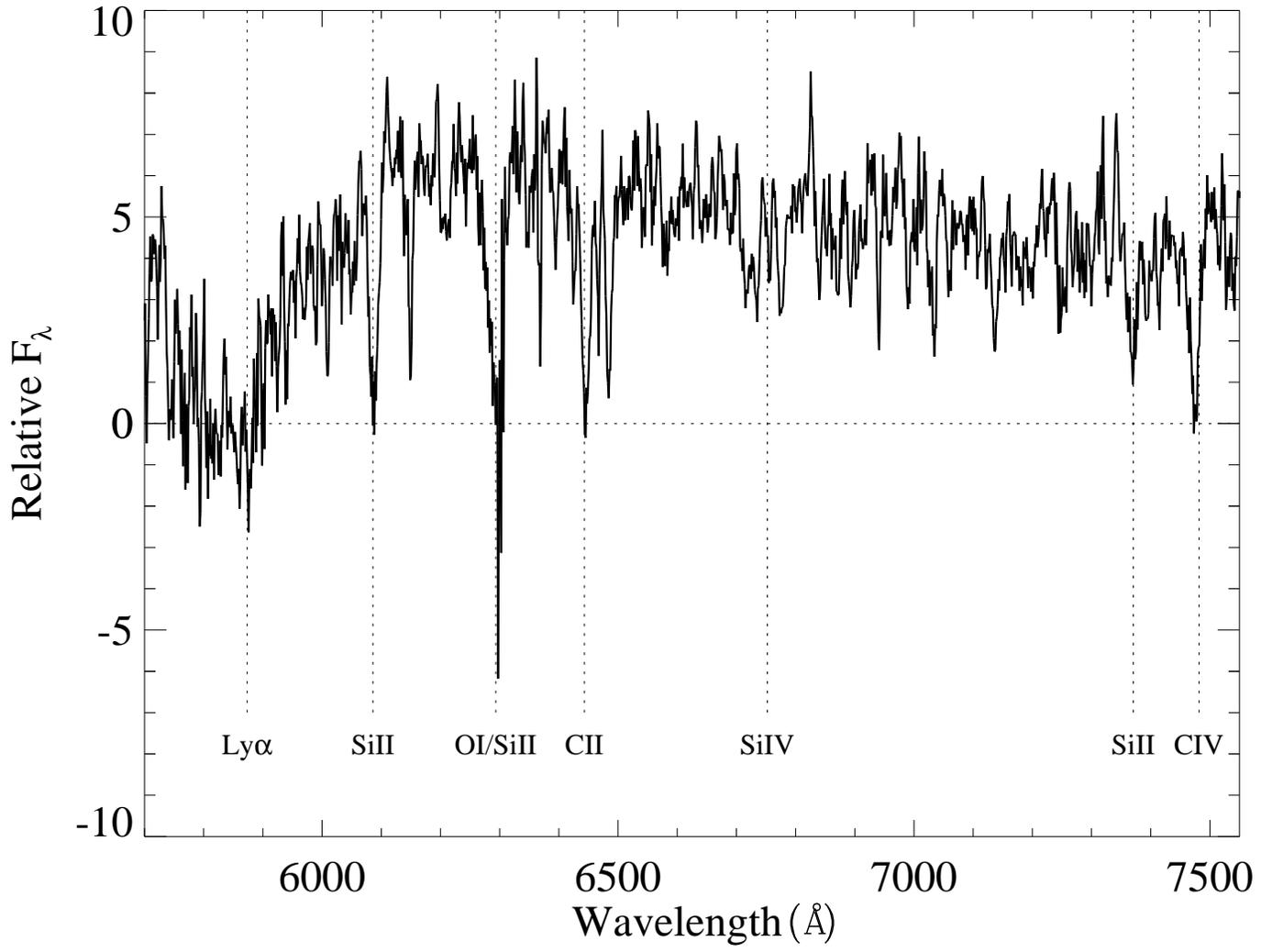}
\end{center}
\caption{\label{fig:spectrum}VLT/FORS2 spectrum of object \#367 with a redshift of $3.830\pm0.002$.}
\end{figure}

We have computed the photometric redshifts of the \gp-dropouts in
TN1338 and the `simulated' \gp-dropouts in GOODS. We let BPZ output
the full redshift probability distribution for each object,
$P_{i}(z)$, and summed over all the objects to get the total redshift
probability distribution. In this manner, information on the
likelihoods of all redshifts are being retained, therby improving the
$S/N$ of the total photometric redshift distribution. The result is
shown in Fig. \ref{fig:bpzprobs}. The area under the curves is equal
to the total number of objects found in a 11.7 arcmin$^2$
area. According to BPZ the fraction of galaxies at $z\sim0.5$ totals
$\sim4$\% of the candidate $z\sim4$ sample in GOODS. The true
contamination fraction is likely to be somewhat higher
\citep{giavalisco04_results}.  The difference in the areas under the
two curves reflects the factor $\sim2.5$ overdensity of the TN1338
field. The peak of the $z_B$ distribution lies at $z=4.1$, which is a
good match to our target redshift as defined by the radio galaxy and
the LAEs. The photometric redshift distribution is steeper around
$z\approx4.1$ for the \gp-dropouts in TN1338 than for those in
GOODS. The narrowness of the distribution and the overdensity is
illustrated by subtracting the GOODS $z_B$ distribution from that of
TN1338 (black curve in Fig. \ref{fig:bpzprobs}).  We conclude that the
field of TN1338 is overdense compared to GOODS at the approximate
redshift of $z\sim4.1$, but structure membership for individual
galaxies is difficult to determine given the relative breadth of the
redshift distribution of $\delta(z)\sim0.5$ (FWHM) based on the
present data.  We note that $N(z)$ has a small secondary peak around
$z\approx3.8$. Interestingly, its redshift corresponds to the other
$\sim6L^*$ object in the field (\#367 at $z=3.83$,
Fig. \ref{fig:spectrum}).  The TN1338 field is exceptional in the
sense that it contains two $\sim6L^*$ LBGs, while there are only a few
of such objects in the entire GOODS field.  \lya\ spectroscopy at this
redshift would be needed to determine whether the TN1338 field
contains another, overlapping, large-scale structure associated with
this object as well.

\section{Summary and discussion}
\label{sec:discussion}

\subsection{The physical properties of LBGs and LAEs}

{\it SFRs} -- We studied the star forming properties of 66
\gp-dropouts to \zp$=27$ mag, and 12 LAEs (6 of which are also in the
\gp-dropout sample). The SFRs were in the range 1--100 M$_\odot$
yr$^{-1}$ (Table \ref{tab:lbgs}), with LAEs being limited to $<14$
M$_\odot$ yr$^{-1}$ (Table \ref{tab:laes}). Applying an average
extinction ($E(B-V)=0.1$ mag) gives SFRs of up to $\sim300$ M$_\odot$
yr$^{-1}$.

{\it Ages/dust} -- The LBGs and LAEs have very blue continua
($\beta_{iz}\approx-2$) when averaged over the entire sample. We
derived a UV-slope magnitude relation of
$\beta_{iz}=(-0.16\pm0.05)(z_{850}-25~\mathrm{mag})-(1.84\pm0.11)$,
and found that the slopes of the $L\gtrsim L^*$ LBGs are consistent
with the average slopes determined for mildly reddened LBGs at
$z\sim3$ redshifted to $z=4$
\citep[Fig. \ref{fig:lbgsbetas};][]{papovich01,shapley03}.  The
$\beta_{iz}-z_{850}$ relation can be interpreted as a SFR- or
mass-extinction relation implying $E(B-V)\approx0.13$ mag at
\zp$\approx23$ mag and $E(B-V)\approx0.0$ mag at \zp$\approx27$ mag
for a fixed age of $\sim70$ Myr.

We derived rest-frame UV-optical colors, and found LBG ages in the
range 10--100 Myr, with $\sim50$\% of the LBGs having ages $<50$ Myr
with respect to our base template \citep[exponentially declining,
$\tau=10$ Myr, $Z=0.2Z_\odot$ and $E(B-V)=0.16$ mag
from][]{papovich01}. We also found evidence for a relation in
\ip--\ks\ vs. \ks, similar as found for \bp-dropouts in GOODS
\citep[Fig. \ref{fig:uvopt1};][]{papovich04}. This is likely to be
interpreted as a stellar mass-age or mass-dust relation, in the sense
that more massive galaxies have higher optical luminosities and redder
UV-optical colors due to aging or dust.

{\it Masses} -- None of the LAEs was detected in the \ks-band, but we
found a $3\sigma$ detection through stacking. The stacked magnitude of
\ks$=25.8^{+0.44}_{-0.32}$ mag implied \ip--\ks$\approx0.0$, while a
stack of LBGs with similar UV magnitudes gave a $5\sigma$ detection of
\ks$=25.14^{+0.23}_{-0.18}$ mag and \ip--\ks$\approx0.7$. Although the
difference in \ks\ magnitude is only $\sim2\sigma$, other studies
based on longer wavelength data suggest that LAEs may indeed be
fainter at optical and infrared wavelengths than LBGs while having
similar SFRs \citep{pentericci07}. The difference can be interpreted
as the LAEs being younger or less massive than LBGs. We use the
\ks-band magnitudes to derive stellar masses, finding
$\sim3\times10^8$ M$_\odot$ for the faintest LAEs and LBGs to
$\sim2\times10^{10}$ M$_\odot$ for the brightest LBGs
(Fig. \ref{fig:uvopt2}). The mean stellar mass of the LAEs in the
protocluster region is in good agreement with the mean stellar masses
of LAEs selected at $z=3-5$ \citep{gawiser06,pirzkal06,pentericci07,nilsson07}.

The possible difference between the LAEs and LBGs is qualitatively
consistent with \citet{charlot93}, who find that \lya\ emission may
primarily escape during a relatively short, dustless phase after the
onset of star formation. However, observations in \lya\ and the UV of
some local starbursts having properties similar to that of high
redshift LBGs seem to suggest instead that the escape of \lya\ photons
may have more to do with the properties of the gas kinematics rather
than with dust \citep{kunth03}.  The nature of the origin of \lya\
emission therefore remains unclear for the moment.

{\it Sizes} -- For $L\ge L^*$ galaxies we found a mean half-light
radius of $0\farcs22$ ($\sim2$ kpc). Although this is slightly smaller
than the average size reported by \citet{ferguson04} based on a GOODS
\bp-dropout sample, \citet{ferguson04} measured $r_{hl}$ out to much
larger circular annuli. The results are consistent when we apply a
small aperture correction from Fig. \ref{fig:completeness}.  The mean
$r_{hl}$ of the \gp-dropouts is comparable to that of \bp-dropouts
culled from the UDF and GOODS fields by \citet{bouwens04_sizes}, and
is therefore consistent with the $\propto(1+z)^{-1.05\pm0.21}$ size
scaling law that connects \up-\bp-,\vp-, and \ip-dropouts at fixed
luminosities \citep{bouwens04_sizes}. The $r_{hl}$ of the $L<L^*$ LAEs
and LBGs are comparable ($\sim1.5$ kpc). The radii of LAEs in the \rp\
filter that contains \lya\ are similar to those in the \ip\ filter
which images our $z\sim4$ sample purely in the UV continuum,
indicating that the highest surface brightness \lya\ originates from
the same region as the stellar continuum.

{\it AGN} -- Although several of the LAEs appear pointlike, the
spectra show no evidence for broad \lya\ or high ionization lines.
Likewise, X-ray observations of a large field sample at $z\sim4.5$
show no positive detections of AGN among \lya\ emitters
\citep{wang04}, and \citet{ouchi07} find an AGN fraction of only
$\simeq1$\% among \lya\ emitters at $z=3-4$.  On the other hand,
several of the LAEs in the protocluster near radio galaxy MRC
1138--262 at $z=2.16$ have been detected with {\it Chandra} indicating
that the AGN fraction of such protoclusters could be significant
\citep[][]{pentericci02,croft05}.

{\it Morphologies} -- The HST images show that the \gp-dropouts span a
wide range of morphologies, with some showing clear evidence for
small-scale interactions (Fig. \ref{fig:lbgzstamps}).  The
morphological parameters $G$, $M_{20}$ and $C$ determined for a bright
subset of the \gp\ dropouts suggest that the morphologies resemble
those of \bp\ dropouts in GOODS \citep[Fig. \ref{fig:nick},][]{lotz04,lotz05}.

\subsubsection{Summary}

\noindent
To summarize, the properties of our UV-selected ``protocluster''
sample of LBGs and LAEs are in good agreement with the physical
properties of galaxies in field samples. In other words, we find no
evidence for trends that can be linked to the relatively rich
environment of the TN1338 structure, in contrast to what has been
observed in lower redshift structures \citep[e.g.][but see
\citet{peter07}]{steidel05,kodama07}.  Even if such trends in stellar mass,
age, size or morphology exist at $z>4$, the relatively crude redshift
selection applied to select the \gp-dropouts at the approximate
redshift of TN1338 ($\delta(z)\sim0.5$, FWHM) may wash out its signal
due to galaxies in the immediate fore- and background.

\subsection{Properties of the protocluster}

\subsubsection{Clustering properties}
\label{sec:halosizes}

\noindent
We have presented evidence for an overdensity of \gp-dropouts in
TN1338 (Fig. \ref{fig:cells}), and that this population has
significant sub-clustering across the ACS field
(Fig. \ref{fig:surfdens}). The radio galaxy lies in a $\sim7.5$
(co-moving) Mpc `filament' formed by the majority of the
\gp-dropouts. The discovery of this substructure with ACS ties in
closely with the clustering seen at larger scales. \citet{intema05}
present the large-scale distribution of relatively bright $B$-dropouts
towards TN1338 in a $25\arcmin\times25\arcmin$\ field observed with
the Subaru Telescope, showing several significant density enhancements
amidst large voids.

In contrast, \citet{venemans02,venemans07}, using data from two
$7\arcmin\times7\arcmin$ VLT/FORS fields, found that the LAEs are
relatively randomly distributed. Based on the absense of substructure
and their small velocity dispersion they concluded that the LAEs might
just be breaking away from the local Hubble flow.  Additional evidence
for this might be contained in the fact that the LAEs in the much
smaller ACS field seem to prefer regions that are generally devoid of
the UV-selected LBGs.  We showed evidence that LAEs are generally
younger (and possibly less massive) than objects in our UV-selected
sample. Also, \gp-dropouts that were detected in \ks\ are both
brighter and redder than LAEs, suggesting that objects with both
higher ages or dust and larger stellar masses lie in the densest
regions of the TN1338 field \citep[see also][]{kashikawa07}. This may
be expected if the oldest and most massive galaxies predominantly end
up in the population of red sequence galaxies dominating the inner
part of clusters at $z\lesssim1.5$.

We have compared the clustering of the \gp-dropouts to that expected
based on the angular clustering of \bp-dropouts at a similar
redshift. We found a small excess of clustering at the smallest
angular scales ($\theta<10\arcsec$) compared to mock samples with a
built-in large-scale correlation function (Fig. \ref{fig:wt}).  The
small-scale excess is to be expected when \wt\ is dominated by
non-linear sub-halo clustering at small scales
\citep{kravtsov04,ouchi05c,lee05}.  A radius of $\sim0.3-0.6$
(co-moving) Mpc is similar to the virial radius, $r_{200}$, of dark
matter halos with masses of $10^{12-13}$ M$_\odot$
\citep[see][]{ouchi05c}, where $r_{200}$ is defined as the radius of a
sphere in which the mean density is $200\times$ the mean density of
the Universe \citep{mo02}. The linear bias, $b$, within these radii
can reach values of $>10-50$, compared to a bias of $2.5-4$ for the
field \citep{ouchi05c,lee05}. Detection of this small-scale clustering
implies that a significant fraction of the \gp-dropouts may share the
same halo, possibly associated with the forming structure near TN1338.

Given their overall strong clustering properties and cosmic variance,
$z\sim4$ LBGs will likely have significant field-to-field variations
even on angular scales larger than currently probed by GOODS
\citep{somerville04}. Future deep, wide surveys will demonstrate the
uniqueness of protocluster structures such as found in the TN1338
field. Based on shallower samples, albeit of significantly larger
areas, it has been found that the surface density of TN1338-like
object concentrations are comparable to that expected based on the
co-moving volume densities of local clusters
\citep[e.g.][]{steidel98,shimasaku03,ouchi05,intema05}. Observations
of these kind of systems could constrain the halo mass function and
the halo occupation distribution at high redshift at the very high
mass end.

\subsubsection{Mass of the overdensity}

\noindent
A proper determination of the mass of the TN1338 protocluster requires
a good estimate of the total volume, which depends both on the angular
and redshift distribution of the LBGs and LAEs. We were not able to
confirm any additional LBGs at the exact redshift of the radio galaxy,
not only because of the faintness and therefore relatively uncertain
photometric redshifts of the targets, but also because all objects
with high equivalent width \lya\ (the most efficient method of
confirming objects at these high redshifts) had already been found
previously in this field.  We nevertheless indirectly obtained
redshifts of the 6 \gp-dropouts that are also in the \lya\ sample
(excluding the radio galaxy).  From the redshift distribution of LBGs
in the GOODS simulations we expect 2.33 field LBGs at a redshift of
$z=4.1$ with $|z-\delta z|<0.03$.  The volume overdensity in TN1338 is
then $\delta_g\equiv(\rho-\bar{\rho})/\bar{\rho}=1.6$, which can be
considered to be a lower limit since it assumes that no other
\gp-dropouts lie within $\delta z$ of $z=4.1$. Alternatively, given
that only 20\%--25\% of field LBGs have large enough equivalent width
\lya\ emission to be detected as a LAE \citep{steidel00}, the 10 LAEs
(with \zp$<27$ mag) associated with TN1338 would represent 40--50 LBGs
(including LAEs) in a relatively narrow redshift range. Note that if
we subtract the average number of \gp-dropouts expected in a
TN1338-sized field ($\sim26$) from the number of \gp-dropouts
observed, the surplus amounts to $\sim40$ \gp-dropouts. This is an
independent confirmation of the results of
\citet{venemans02,venemans07} that were based on LAEs alone.

If we assume that the LBGs in TN1338 have the same overdensity as
measured for the LAEs by \citet{venemans02,venemans07}, the
overdensity in the TN1338 ACS field would be $\delta_g\approx3.5$.  We
can relate the true mass overdensity, $\delta_m$, to the observed
\gp-dropout overdensity, $\delta_g$, through
$1+b\delta_m=|C|(1+\delta_g)$, where $C=1+f-f(1+\delta_m)^{1/3}$ (with
$f=\Omega_m(z)^{4/7}$) corrects for redshift-space distortion due to
peculiar velocities assuming that the structure is just breaking away
from the Hubble expansion \citep{steidel98}. Taking $b\sim3$ for
\bp-dropouts with \zp$<26-27$ from \citet{lee05} gives
$\delta_m\sim0.8$ for $\delta_g=3.5$. This can be related to a total
mass of $M\simeq(1+\delta_m)\bar{\rho}V\gtrsim10^{14}$ M$_\odot$,
where $\bar{\rho}$ is the present-day mean density of the Universe and
$V$ is the volume probed by our ACS observations.  The mass
overdensity corresponds to a linear overdensity of $\delta_L\sim0.5$,
which when evolved to the present epoch corresponds to a linear
overdensity of $\delta_L\sim2$ \citep[see][]{steidel05,overzier06b}.  This exceeds
the linear collapse threshold of $\delta_c=1.69$, so this
``protocluster'' will have virialized by $z=0$.  We note that these
calculations depend on several critical assumptions, such as the size
of the volume \citep[likely to be larger than the ACS
field,][]{venemans02,intema05}, the magnitude of the overdensity (here
assumed to be constant over the volume), and the exact bias value
corresponding to the objects used to determine the overdensity.  The
validity of these assumptions and the typical properties of such
structures will be the subject of a detailed comparison with numerical
simulations in a future work (Overzier et al., in prep.).

\subsubsection{Redshift evolution of the overdensity} 

\noindent
The progenitors of galaxy clusters must have undergone rapid and
intense star-formation (and possibly AGN activity) at $z\gtrsim2$.
The star formation in these `protoclusters' is not only responsible
for the buildup of the present-day stellar mass in cluster galaxies,
but also for the chemical enrichment of the intracluster medium
\citep[e.g.][]{ettori05,venemans05}.  The ages of the stellar
populations in massive red-sequence galaxies at $z\sim1$ are
sufficiently high for them to have begun forming at redshifts
$2\lesssim z\lesssim5$ \citep[e.g.][and references
therein]{blakeslee03,glazebrook04,mei06,homeier06,rettura06}.

Simulations suggest that there may be significant differences between
the redshift of formation and redshift of assembly for the stellar
mass in massive early-type galaxies: \citet{delucia05} found that for
elliptical galaxies with stellar mass larger than $10^{11}$ M$_\odot$
the median redshift at which 50\% of the stars were formed is
$\sim2.5$, but the median redshift when those stars were actually
assembled into a single galaxy lies only at $\sim0.8$.  In the same
paper, they showed that the star formation properties of ellipticals
depend strongly on the environment. For elliptical galaxies in
clusters, the average ages can be up to 2 Gyr higher than those of
similar mass ellipticals in the field. For an elliptical that is
$\sim1$ Gyr older compared to the field, 50\% of its stellar mass will
already have been formed at $z\sim4$. This stellar mass is likely to
be formed in much smaller units, while the number of major mergers is
considered to be relatively small (a few).

It is likely that the stellar mass formed by protocluster galaxies
will end up in quiescent cluster early-types at lower redshifts, in
accordance with the color-magnitude and morphology-density
relations. However, there is a significant discrepancy between the
masses of the LBGs and LAEs (both in protoclusters and in the field)
and the masses of cluster early-type galaxies of $\gtrsim10$,
indicating that a large fraction of the stellar mass still has to
accumulate through merging. Detailed observations of protocluster
regions on much larger scales ($\sim$50 co-moving Mpc) are needed to
test if the number density of LBGs is indeed consistent with forming
the cluster red sequence population through merging.

Alternatively, protocluster fields may also host older galaxies of
significant mass, analogous to the population of distant red galaxies
found at $2<z<4$ \citep[e.g.][]{franx03,vandokkum03,webb05}.  These
objects are believed to be the aged and reddened descendants of LBGs
that were UV luminous only at $z\gtrsim5-6$, and form a population
that is highly clustered \citep{quadri07}. Attemps are currently being
made in finding such objects toward protoclusters
\citep[e.g.][]{kajisawa06,kodama07}, but spectroscopic confirmation is
difficult.

\subsection{Comparison to simulations}

\noindent
Our results are in agreement with studies of large-scale structure and protoclusters 
using $N$-body simulations. \citet{suwa06} studied  
global properties of protoclusters by picking up the dark matter particles belonging to clusters at $z=0$ 
and tracing them back to high redshift. 
The simulations showed that clusters with masses of $>10^{14}$ $M_\odot$ can be traced back to 
regions at $z=4-5$ of 20--40 $h^{-1}$ Mpc in size, and that these regions 
are associated with overdensities of typical halos hosting LAEs and LBGs 
of $\delta_g\sim3$ and mass overdensities $\delta_m$ in the range 0.2--0.6. 
For randomly selected regions of the same size, the galaxy and mass overdensities were 
found to be mostly $\lesssim0$, as expected due to the fact that massive halos are relatively rare. Although 
some of the overdense regions in the simulations having a similar overdensity as our protocluster 
candidate do not end up in clusters at $z=0$, the simulations show that most regions with an 
overdensity on the order of a few at $z\sim5$ will evolve into clusters more massive than $10^{14}$ 
$M_\odot$ ($\gtrsim50$\% for $\delta_g\gtrsim2$). 

Also, recent numerical simulations of CDM growth predict that quasars at $z\sim6$ may lie in the 
center of very massive dark matter halos of $\sim4\times10^{12}$ $M_\odot$ \citep{springel05,li06}. 
They are surrounded by many fainter galaxies, that will evolve into massive 
clusters of $\sim 4\times10^{15}$ $M_\odot$ at $z=0$.  
The discovery of galaxy clustering associated with luminous radio galaxies 
and quasars at $z>2$ \citep[e.g.][this paper]{stiavelli05,venemans07,zheng06} 
is consistent with that scenario. 

\begin{acknowledgements}
We thank Masami Ouchi for invaluable discussions and reading through the manuscript. 
We would also like to thank Ryan Quadri and Huib Intema for their contributions, 
and the anonymous referee for his or her suggestions. 

ACS was developed under NASA contract NAS 5-32865, and this research
has been supported by NASA grant NAG5-7697 and
by an equipment grant from Sun Microsystems, Inc.
The {Space Telescope Science Institute} is operated by AURA Inc., under
NASA contract NAS5-26555. We are grateful to K.~Anderson, J.~McCann,
S.~Busching, A.~Framarini, S.~Barkhouser, and T.~Allen for their
invaluable contributions to the ACS project at JHU. JK was supported by DFG grant SFB-439.
\end{acknowledgements}

\begin{deluxetable}{llrlcc} 
\tabletypesize{\scriptsize}
\tablecolumns{6} 
\tablewidth{0pc} 
\tablecaption{\label{tab:log}Summary of observations.} 
\tablehead{\multicolumn{1}{c}{Filter} & \multicolumn{1}{c}{Date} & \multicolumn{1}{c}{$T_{exp}$} & \multicolumn{1}{c}{$A$} & \multicolumn{2}{c}{Depth}\\
\colhead{}     & \colhead{}        &  \multicolumn{1}{c}{(s)}  &  \multicolumn{1}{c}{(mag)} &  \multicolumn{1}{c}{($2\sigma$)} & \multicolumn{1}{c}{($5\sigma$)}}
\startdata 
\gp\ (F475W)  & 2002 July 11--12 &   9400     & 0.359   & 28.46$^a$ & 27.47$^a$\\
\rp\ (F625W)  & 2002 July 8--9 &   9400     & 0.256   & 28.23$^a$ & 27.23$^a$\\
\ip\ (F775W)  & 2002 July 8--9 &  11700     & 0.193   & 28.07$^a$ & 27.08$^a$\\
\zp\ (F850LP) & 2003 July 11--12 &  11800     & 0.141   & 27.73$^a$ & 26.73$^a$\\
$K_S$         & 2002 March 24--26, 2004 May--July   &  27000     & 0.036   & 25.15$^b$ & 24.16$^b$\\
\enddata 
\tablenotetext{a}{Measured in 0\farcs45 diameter square apertures.}
\tablenotetext{b}{Measured in 1\farcs4 diameter circular apertures.}
\end{deluxetable} 

\begin{deluxetable}{llllrrrrllr}
\tabletypesize{\scriptsize}
\tablecolumns{11} 
\tablewidth{0pc} 
\tablecaption{\label{tab:laes}Properties of the spectroscopically confirmed Ly$\alpha$\ emitters.}
\tablehead{\multicolumn{1}{c}{ID} & \multicolumn{1}{c}{$\alpha_{J2000}$} & \multicolumn{1}{c}{$\delta_{J2000}$} & \multicolumn{1}{c}{$z_{spec}$} & \multicolumn{1}{c}{(\gp--\rp)$^a$} & \multicolumn{1}{c}{(\rp--\zp)$^a$} & \multicolumn{1}{c}{(\ip--\zp)$^a$} & \multicolumn{1}{c}{\zp$^b$} & \multicolumn{1}{c}{$r_{hl,r}$} & \multicolumn{1}{c}{$r_{hl,z}$} & \multicolumn{1}{c}{$SFR_{UV}^c$}}
\startdata 
RG & 13:38:26.05 & --19:42:30.47 & 4.105 &$3.42\pm0.17$ & $-0.61\pm0.03$ & $0.09\pm0.03$ & $23.05\pm0.05$ &  $0\farcs60$& $0\farcs62$ & $93.7^{+2.71}_{-2.63}$\\  
L4 & 13:38:22.46 & --19:44:33.67 & 4.095 &$1.88\pm0.27$ & $-0.35\pm0.15$ & $-0.14\pm0.16$ & $26.68\pm0.23$ & $0\farcs09$ & $0\farcs08$ &  $4.83^{+0.59}_{-0.53}$\\
L7 & 13:38:24.78 & --19:41:33.66 & 4.106 &$>1.26$ & $0.65\pm0.29$ & $0.32\pm0.25$ & $27.20\pm0.49$ & $0\farcs18$ & $0\farcs11$ &  $3.22^{+0.85}_{-0.67}$\\
L8 & 13:38:24.86 & --19:41:45.49 & 4.102 &$>1.47$ & $0.37\pm0.27$ & $-0.26\pm0.23$ & $26.51\pm0.30$ & $0\farcs12$ & $0\farcs16$ &  $6.25^{+0.88}_{-0.77}$\\
L9 & 13:38:25.10 & --19:43:10.77 & 4.100 &$1.78\pm0.25$ & $0.49\pm0.09$ & $-0.02\pm0.07$ & $25.34\pm0.08$ & $0\farcs12$ & $0\farcs14$ &  $14.4^{+0.60}_{-0.58}$\\
L11 & 13:38:26.16 & --19:43:34.31 & 4.101 & $1.56\pm0.18$ & $0.11\pm0.09$ & $-0.08\pm0.09$ & $25.94\pm0.10$ & $0\farcs08$ & $0\farcs09$ &  $8.95^{+0.48}_{-0.46}$\\
L14 & 13:38:28.72 & --19:44:36.98 & 4.102 &$>1.88$ & $0.37\pm0.18$ & $0.24\pm0.18$ & $26.52\pm0.18$ & $0\farcs13$ & $0\farcs13$ &  $4.29^{+0.54}_{-0.48}$\\
L16 & 13:38:29.66 & --19:43:59.82 & 4.102 &$1.43\pm0.23$ & $0.22\pm0.12$ & $-0.00\pm0.11$ & $25.54\pm0.16$ & $0\farcs16$ & $0\farcs19$ &  $11.4^{+1.10}_{-1.00}$\\
L17 & 13:38:29.86 & --19:43:25.84 & 4.093 &$>1.57$ & $0.21\pm0.27$ & $0.44\pm0.29$ & $27.37\pm0.28$ & $0\farcs11$ & $0\farcs10$ &  $1.33^{+0.40}_{-0.31}$\\
L20 & 13:38:32.83 & --19:44:6.934 & 4.100 &$>2.17$ & $0.37\pm0.15$ & $0.13\pm0.14$ & $26.44\pm0.16$ & $0\farcs09$ & $0\farcs11$ &  $5.12^{+0.53}_{-0.48}$\\
L21 & 13:38:33.56 & --19:43:36.00 & 4.097 &$1.50\pm0.32$ & $-0.00\pm0.18$ & $-0.16\pm0.18$ & $26.22\pm0.19$ & $0\farcs12$ & $0\farcs18$ &  $4.75^{+0.79}_{-0.68}$\\
L22 & 13:38:34.14 & --19:42:52.68 & 4.096 &$>1.82$ & $0.79\pm0.16$ & $0.13\pm0.13$ & $26.64\pm0.14$ & $0\farcs07$ & $0\farcs08$ &  $4.69^{+0.38}_{-0.35}$\\
L25 & 13:38:34.96 & --19:42:24.95 & 4.093 &$1.68\pm0.33$ & $0.32\pm0.13$ & $0.00\pm0.12$ & $25.81\pm0.15$ & $0\farcs25$ & $0\farcs24$ &  $9.94^{+0.78}_{-0.72}$\\
\enddata 
\tablenotetext{a}{Isophotal colors. The limits are $2\sigma$.}
\tablenotetext{b}{Total magnitudes.}
\tablenotetext{c}{SFR estimated from the UV continuum flux (\ip).}
\end{deluxetable} 

\begin{deluxetable}{llllrrlll}
\tabletypesize{\tiny}
\tablecolumns{9} 
\tablewidth{0pc} 
\tablecaption{\label{tab:lbgs}Properties of the $z\sim4$ Lyman break sample.}
\tablehead{\multicolumn{1}{c}{ID} & \multicolumn{1}{c}{$\alpha_{J2000}$} & \multicolumn{1}{c}{$\delta_{J2000}$}  & \multicolumn{1}{c}{(\gp--\rp)$^a$} & \multicolumn{1}{c}{(\rp--\zp)$^a$} & \multicolumn{1}{c}{(\ip--\zp)$^a$} & \multicolumn{1}{c}{\zp$^b$} & \multicolumn{1}{c}{$r_{hl,z}$} & \multicolumn{1}{c}{$SFR_{UV}^c$}}
\startdata 
2707/RG & 13:38:26.05 & --19:42:30.47 & $3.42\pm0.17$ & $-0.61\pm0.03$ & $0.09\pm0.03$ & $23.05\pm0.05$ & $0\farcs62$ & $93.7^{+2.71}_{-2.63}$\\           
367     &13:38:32.75 & --19:44:37.27 & $1.70\pm0.06$ & $0.52\pm0.02$ & $0.07\pm0.02$ & $23.10\pm0.02$ & $0\farcs20$ & $94.7^{+1.10}_{-1.09}$\\             
1991    & 13:38:27.84 & --19:43:15.19 & $1.88\pm0.25$ & $0.43\pm0.08$ & $0.05\pm0.07$ & $24.43\pm0.12$ & $0\farcs42$ & $28.8^{+2.24}_{-2.08}$\\            
3018    & 13:38:24.31 & --19:42:58.06 & $1.73\pm0.14$ & $0.29\pm0.06$ & $-0.01\pm0.05$ & $24.49\pm0.07$ & $0\farcs20$ & $34.2^{+1.19}_{-1.15}$\\           
3216    & 13:38:22.37 & --19:43:32.41 & $1.86\pm0.15$ & $0.33\pm0.05$ & $0.10\pm0.05$ & $24.54\pm0.06$ & $0\farcs21$ & $26.3^{+0.92}_{-0.89}$\\            
3116    & 13:38:24.21 & --19:42:41.85 & $1.55\pm0.14$ & $0.36\pm0.06$ & $-0.00\pm0.06$ & $24.67\pm0.06$ & $0\farcs20$ & $25.4^{+0.93}_{-0.90}$\\           
959     &13:38:32.67 & --19:43:3.673 & $1.88\pm0.25$ & $0.55\pm0.07$ & $0.28\pm0.07$ & $24.73\pm0.09$ & $0\farcs23$ & $18.2^{+1.19}_{-1.12}$\\             
2913    & 13:38:23.68 & --19:43:36.59 & $1.77\pm0.22$ & $0.47\pm0.07$ & $-0.01\pm0.06$ & $24.94\pm0.10$ & $0\farcs26$ & $24.2^{+1.16}_{-1.11}$\\           
2152    & 13:38:26.92 & --19:43:27.60 & $1.56\pm0.17$ & $0.15\pm0.08$ & $-0.01\pm0.08$ & $24.97\pm0.15$ & $0\farcs32$ & $22.4^{+1.74}_{-1.62}$\\           
2799    & 13:38:24.88 & --19:43:7.415 & $1.72\pm0.17$ & $0.19\pm0.07$ & $-0.03\pm0.07$ & $25.03\pm0.09$ & $0\farcs18$ & $18.6^{+0.97}_{-0.92}$\\           
2439    & 13:38:25.35 & --19:43:43.65 & $1.69\pm0.24$ & $0.46\pm0.09$ & $0.14\pm0.08$ & $25.08\pm0.10$ & $0\farcs25$ & $17.7^{+1.03}_{-0.97}$\\            
3430    & 13:38:21.21 & --19:43:41.99 & $1.74\pm0.22$ & $0.49\pm0.08$ & $-0.03\pm0.07$ & $25.10\pm0.09$ & $0\farcs16$ & $17.0^{+0.88}_{-0.83}$\\           
2407    & 13:38:24.35 & --19:44:29.15 & $1.54\pm0.21$ & $0.43\pm0.09$ & $0.04\pm0.08$ & $25.11\pm0.11$ & $0\farcs26$ & $15.6^{+1.04}_{-0.97}$\\            
2839    & 13:38:25.90 & --19:42:18.39 & $2.23\pm0.45$ & $0.75\pm0.09$ & $0.12\pm0.07$ & $25.14\pm0.10$ & $0\farcs15$ & $15.6^{+0.98}_{-0.92}$\\            
1252    & 13:38:31.98 & --19:42:37.47 & $1.57\pm0.16$ & $0.39\pm0.07$ & $-0.03\pm0.06$ & $25.25\pm0.08$ & $0\farcs11$ & $16.3^{+0.70}_{-0.67}$\\           
227     &13:38:33.02 & --19:44:47.57 & $1.57\pm0.17$ & $-0.06\pm0.09$ & $-0.14\pm0.09$ & $25.29\pm0.19$ & $0\farcs23$ & $20.4^{+1.22}_{-1.15}$\\           
2710/L9 & 13:38:25.10 & --19:43:10.77 & $1.78\pm0.25$ & $0.49\pm0.09$ & $-0.02\pm0.07$ & $25.34\pm0.08$ & $0\farcs14$ & $14.4^{+0.60}_{-0.58}$\\           
1815    & 13:38:29.01 & --19:43:3.275 & $1.67\pm0.24$ & $0.27\pm0.11$ & $0.01\pm0.10$ & $25.50\pm0.12$ & $0\farcs17$ & $11.8^{+0.80}_{-0.75}$\\            
1152    & 13:38:32.62 & --19:42:25.15 & $>2.13$ & $0.48\pm0.14$ & $-0.19\pm0.11$ & $25.57\pm0.20$ & $0\farcs28$ & $15.0^{+1.28}_{-1.18}$\\                         
2755    & 13:38:24.95 & --19:43:16.89 & $1.78\pm0.53$ & $0.36\pm0.19$ & $0.12\pm0.18$ & $25.59\pm0.21$ & $0\farcs33$ & $10.3^{+1.40}_{-1.23}$\\            
3304    & 13:38:23.67 & --19:42:27.37 & $1.66\pm0.33$ & $0.56\pm0.12$ & $0.08\pm0.10$ & $25.60\pm0.12$ & $0\farcs15$ & $10.2^{+0.77}_{-0.71}$\\            
1819    & 13:38:29.61 & --19:42:38.19 & $1.98\pm0.33$ & $0.39\pm0.10$ & $0.15\pm0.09$ & $25.60\pm0.15$ & $0\farcs14$ & $11.0^{+0.97}_{-0.89}$\\            
3159    & 13:38:22.21 & --19:43:50.13 & $>1.59$ & $0.49\pm0.23$ & $0.11\pm0.20$ & $25.63\pm0.16$ & $0\farcs41$ & $7.83^{+1.04}_{-0.92}$\\                  
309     &13:38:34.77 & --19:43:27.59 & $1.55\pm0.27$ & $0.20\pm0.13$ & $0.06\pm0.12$ & $25.69\pm0.15$ & $0\farcs21$ & $10.1^{+0.92}_{-0.85}$\\             
1808    & 13:38:30.04 & --19:42:22.51 & $1.61\pm0.27$ & $0.36\pm0.11$ & $-0.10\pm0.10$ & $25.71\pm0.12$ & $0\farcs15$ & $10.1^{+0.73}_{-0.68}$\\           
3670    & 13:38:20.73 & --19:43:16.32 & $2.09\pm0.47$ & $0.50\pm0.12$ & $0.07\pm0.10$ & $25.81\pm0.11$ & $0\farcs14$ & $8.77^{+0.61}_{-0.57}$\\            
633/L25 &13:38:34.96 & --19:42:24.95 & $1.68\pm0.33$ & $0.32\pm0.13$ & $0.00\pm0.12$ & $25.81\pm0.15$ & $0\farcs24$ & $9.94^{+0.78}_{-0.72}$\\             
2524    & 13:38:24.47 & --19:44:7.263 & $>2.01$ & $0.60\pm0.14$ & $-0.02\pm0.12$ & $25.86\pm0.21$ & $0\farcs17$ & $13.3^{+1.03}_{-0.96}$\\                         
1461    & 13:38:31.37 & --19:42:30.95 & $>1.97$ & $0.56\pm0.15$ & $0.07\pm0.13$ & $25.89\pm0.15$ & $0\farcs20$ & $8.22^{+0.77}_{-0.70}$\\                  
3177    & 13:38:22.97 & --19:43:16.07 & $1.61\pm0.35$ & $0.27\pm0.16$ & $0.12\pm0.15$ & $25.99\pm0.23$ & $0\farcs18$ & $8.70^{+1.09}_{-0.97}$\\            
1668    & 13:38:26.93 & --19:44:53.25 & $>1.82$ & $0.72\pm0.17$ & $0.16\pm0.13$ & $25.99\pm0.16$ & $0\farcs17$ & $5.58^{+0.73}_{-0.64}$\\                  
358     &13:38:32.12 & --19:45:4.687 & $2.00\pm0.54$ & $0.48\pm0.15$ & $0.03\pm0.13$ & $25.99\pm0.19$ & $0\farcs20$ & $7.87^{+0.87}_{-0.79}$\\             
2569    & 13:38:26.38 & --19:42:43.55 & $1.94\pm0.44$ & $0.36\pm0.14$ & $0.03\pm0.13$ & $26.04\pm0.15$ & $0\farcs13$ & $7.08^{+0.64}_{-0.59}$\\            
3131    & 13:38:25.27 & --19:41:55.49 & $>2.15$ & $0.28\pm0.15$ & $-0.09\pm0.14$ & $26.04\pm0.19$ & $0\farcs16$ & $8.29^{+0.83}_{-0.75}$\\                         
2527    & 13:38:27.99 & --19:41:44.07 & $1.79\pm0.29$ & $0.28\pm0.11$ & $-0.13\pm0.10$ & $26.11\pm0.22$ & $0\farcs07$ & $10.8^{+0.89}_{-0.83}$\\           
2347    & 13:38:27.99 & --19:42:12.22 & $1.59\pm0.40$ & $0.42\pm0.17$ & $0.32\pm0.16$ & $26.14\pm0.19$ & $0\farcs14$ & $5.63^{+0.79}_{-0.69}$\\            
2358    & 13:38:24.12 & --19:44:47.21 & $1.70\pm0.42$ & $0.36\pm0.16$ & $-0.12\pm0.15$ & $26.15\pm0.16$ & $0\farcs14$ & $6.28^{+0.66}_{-0.60}$\\           
307     &13:38:32.80 & --19:44:46.47 & $2.02\pm0.49$ & $0.30\pm0.15$ & $-0.04\pm0.13$ & $26.17\pm0.22$ & $0\farcs15$ & $9.76^{+0.87}_{-0.80}$\\            
2989    & 13:38:22.90 & --19:43:59.01 & $1.93\pm0.29$ & $-0.17\pm0.13$ & $-0.22\pm0.13$ & $26.21\pm0.15$ & $0\farcs12$ & $6.61^{+0.59}_{-0.54}$\\          
552/L21 &13:38:33.56 & --19:43:36.00 & $1.50\pm0.32$ & $-0.00\pm0.18$ & $-0.16\pm0.18$ & $26.22\pm0.19$ & $0\farcs18$ & $4.75^{+0.79}_{-0.68}$\\           
507     &13:38:34.26 & --19:43:12.20 & $>1.92$ & $0.78\pm0.15$ & $0.10\pm0.11$ & $26.22\pm0.12$ & $0\farcs12$ & $6.50^{+0.47}_{-0.44}$\\                   
3564    & 13:38:23.34 & --19:41:51.44 & $1.82\pm0.49$ & $-0.05\pm0.23$ & $-0.02\pm0.23$ & $26.22\pm0.29$ & $0\farcs32$ & $6.64^{+1.11}_{-0.95}$\\          
540     &13:38:33.26 & --19:43:49.45 & $1.85\pm0.52$ & $-0.07\pm0.24$ & $-0.08\pm0.24$ & $26.28\pm0.26$ & $0\farcs23$ & $5.65^{+0.94}_{-0.80}$\\           
2480    & 13:38:27.25 & --19:42:30.43 & $1.78\pm0.40$ & $0.26\pm0.15$ & $-0.04\pm0.14$ & $26.30\pm0.21$ & $0\farcs11$ & $8.26^{+0.75}_{-0.69}$\\           
2712    & 13:38:26.54 & --19:42:12.01 & $1.73\pm0.43$ & $0.48\pm0.15$ & $-0.10\pm0.13$ & $26.34\pm0.18$ & $0\farcs12$ & $7.77^{+0.60}_{-0.56}$\\           
2494    & 13:38:25.39 & --19:43:34.79 & $1.61\pm0.33$ & $0.32\pm0.15$ & $-0.14\pm0.13$ & $26.35\pm0.19$ & $0\farcs09$ & $6.55^{+0.66}_{-0.60}$\\           
2708    & 13:38:24.13 & --19:43:50.55 & $>1.94$ & $-0.03\pm0.23$ & $-0.04\pm0.23$ & $26.39\pm0.22$ & $0\farcs23$ & $3.89^{+0.75}_{-0.63}$\\                        
538/L20 &13:38:32.83 & --19:44:6.934 & $>2.17$ & $0.37\pm0.15$ & $0.13\pm0.14$ & $26.44\pm0.16$ & $0\farcs11$ & $5.12^{+0.53}_{-0.48}$\\                   
1843    & 13:38:29.54 & --19:42:38.83 & $1.82\pm0.44$ & $0.11\pm0.18$ & $-0.21\pm0.17$ & $26.47\pm0.21$ & $0\farcs12$ & $5.44^{+0.65}_{-0.58}$\\           
1876    & 13:38:30.04 & --19:42:27.78 & $>1.61$ & $0.40\pm0.23$ & $0.03\pm0.21$ & $26.48\pm0.25$ & $0\farcs17$ & $5.05^{+0.82}_{-0.71}$\\                  
375     &13:38:32.71 & --19:44:38.30 & $>1.58$ & $0.35\pm0.25$ & $0.09\pm0.23$ & $26.49\pm0.27$ & $0\farcs19$ & $2.93^{+0.89}_{-0.68}$\\                   
1655    & 13:38:29.52 & --19:43:10.60 & $>2.02$ & $0.46\pm0.15$ & $0.27\pm0.14$ & $26.51\pm0.18$ & $0\farcs10$ & $4.69^{+0.55}_{-0.50}$\\                  
1339/L14& 13:38:28.72 & --19:44:36.98 & $>1.88$ & $0.37\pm0.18$ & $0.24\pm0.18$ & $26.52\pm0.18$ & $0\farcs13$ & $4.29^{+0.54}_{-0.48}$\\                  
286     &13:38:34.08 & --19:43:58.08 & $1.63\pm0.45$ & $0.47\pm0.17$ & $-0.20\pm0.14$ & $26.52\pm0.19$ & $0\farcs10$ & $5.43^{+0.55}_{-0.50}$\\            
3133    & 13:38:23.75 & --19:42:56.64 & $>1.95$ & $0.16\pm0.20$ & $0.04\pm0.19$ & $26.53\pm0.27$ & $0\farcs14$ & $5.95^{+0.80}_{-0.70}$\\                  
1800    & 13:38:29.65 & --19:42:39.80 & $>1.92$ & $0.16\pm0.20$ & $-0.01\pm0.19$ & $26.56\pm0.26$ & $0\farcs14$ & $3.72^{+0.79}_{-0.65}$\\                         
3486    & 13:38:21.49 & --19:43:21.68 & $1.82\pm0.45$ & $0.23\pm0.17$ & $-0.06\pm0.16$ & $26.61\pm0.21$ & $0\farcs10$ & $4.84^{+0.59}_{-0.53}$\\           
2874/L4 & 13:38:22.46 & --19:44:33.67 & $1.88\pm0.27$ & $-0.35\pm0.15$ & $-0.14\pm0.16$ & $26.68\pm0.23$ & $0\farcs08$ & $4.83^{+0.59}_{-0.53}$\\          
1211    & 13:38:33.53 & --19:42:9.188 & $>1.73$ & $0.21\pm0.24$ & $0.19\pm0.24$ & $26.72\pm0.27$ & $0\farcs13$ & $2.39^{+0.73}_{-0.56}$\\                  
1203    & 13:38:31.76 & --19:42:53.82 & $1.72\pm0.45$ & $0.42\pm0.17$ & $0.04\pm0.15$ & $26.73\pm0.26$ & $0\farcs07$ & $4.20^{+0.66}_{-0.57}$\\            
2571    & 13:38:23.70 & --19:44:32.20 & $>1.75$ & $0.09\pm0.25$ & $-0.00\pm0.25$ & $26.76\pm0.23$ & $0\farcs14$ & $1.63^{+0.59}_{-0.43}$\\                         
1265    & 13:38:28.64 & --19:44:52.16 & $>2.08$ & $0.22\pm0.18$ & $0.36\pm0.19$ & $26.82\pm0.18$ & $0\farcs11$ & $2.38^{+0.41}_{-0.35}$\\                  
1712    & 13:38:27.93 & --19:44:5.602 & $1.57\pm0.25$ & $-0.10\pm0.14$ & $-0.26\pm0.14$ & $26.83\pm0.31$ & $0\farcs09$ & $6.78^{+0.66}_{-0.60}$\\          
3013    & 13:38:24.18 & --19:43:3.364 & $1.54\pm0.42$ & $0.18\pm0.21$ & $0.05\pm0.21$ & $26.85\pm0.30$ & $0\farcs12$ & $3.79^{+0.69}_{-0.59}$\\            
1866    & 13:38:29.33 & --19:42:44.09 & $>1.63$ & $0.19\pm0.26$ & $0.24\pm0.27$ & $26.86\pm0.30$ & $0\farcs14$ & $2.75^{+0.69}_{-0.55}$\\                  
1290    & 13:38:32.15 & --19:42:25.57 & $>1.54$ & $0.24\pm0.26$ & $-0.05\pm0.24$ & $26.88\pm0.29$ & $0\farcs17$ & $4.00^{+0.66}_{-0.56}$\\                      
\enddata 
\tablenotetext{a}{Isophotal colors. The limits are $2\sigma$.}
\tablenotetext{b}{Total magnitudes.}
\tablenotetext{c}{SFR estimated from the UV continuum flux (\ip).}
\end{deluxetable}

\clearpage

\end{document}